\documentclass[aps,preprint,nofootinbib]{revtex4-1}
\usepackage{amsmath}
\usepackage{latexsym}
\usepackage{nicefrac}
\usepackage{comment}
\usepackage{graphicx}
\usepackage{hyperref}
\usepackage{comment}
\hypersetup{
colorlinks=true,
linkcolor=blue,
citecolor=blue,
urlcolor=blue
}

\usepackage{caption}
\usepackage{subcaption}
\usepackage{geometry}
\usepackage{array}
\usepackage{float}
\usepackage{multirow}
\usepackage{makecell}

\begin{document}

\title{Exact, non-singular black holes from a phantom DBI field as primordial dark matter}
\author{Tausif Parvez}
\email{214120002@iitb.ac.in}

\author{S. Shankaranarayanan}
\email{shanki@iitb.ac.in}

\affiliation{Department of Physics,  Indian Institute of Technology Bombay, Mumbai 400076, India}

\date{\today}


\begin{abstract}
We present the first exact, non-singular black hole solution in General Relativity sourced by a Dirac-Born-Infeld (DBI) scalar field. 
Crucially, the solution is exclusively supported by \emph{the phantom branch of the DBI action}, dynamically replacing the central singularity with a regular core.
The solution is asymptotically flat, possesses non-trivial scalar hair, and replaces the central singularity with a regular 2-sphere. 
The mechanism for singularity resolution is a dynamical 
\emph{kinetic stiffness} which also explains the evasion of classical no-hair theorems. 
%
We show these black holes evaporate to a  non-singular relic with mass of the order of a gram. This provides a robust mechanism to evade standard evaporation constraints, opening a vast, previously forbidden mass window for light \emph{Primordial Black Holes} to constitute dark matter. 
The model is testable via distinctive gravitational-wave signatures from its scalar hair.
\end{abstract}

\maketitle

\section{Introduction}
The inevitability of spacetime singularities, established by the Penrose-Hawking theorems~\cite{1965-Penrose-PRL,Hawking:1970zqf}, remains a central pathology of classical General Relativity (GR)~\cite{Shankaranarayanan:2022wbx}. For the vast majority of known black hole (BH) solutions, both in GR and in various modified gravity theories, the approach to the central singularity at $r=0$ is marked by a catastrophic, power-law divergence in scalar quantities such as the Ricci scalar $R$, the Ricci square $R_{\mu\nu}R^{\mu\nu}$, and the Kretschmann scalar $K = R_{\mu\nu\rho\sigma}R^{\mu\nu\rho\sigma}$~\cite{Mandal:2025xuc}. This near-universal behavior, rigorously classified for a wide class of spacetimes by Szekeres and Iyer~\cite{Szekeres:1993qe,Celerier:2002yi}, underscores the singularity as a fundamental pathology of our current understanding of gravity in the strong-field regime~\cite{Shankaranarayanan:2022wbx,Mandal:2025xuc}.

It is widely understood that resolving these singularities requires a repulsive component to counteract the standard attractive nature of gravity at high curvatures~\cite{1973-Hawking.Ellis-Book,Senovilla:2014gza}. A popular and phenomenologically successful approach has been to introduce this repulsive effect via a mechanism that effectively mimics a positive cosmological constant in the strong-field limit. This leads to the paradigm of \emph{regular BHs}, where the central singularity is excised and replaced by a regular region of spacetime, often a de Sitter core. The first such model was proposed by Bardeen~\cite{Bardeen:1968abc}, and subsequent examples include those by Hayward and others~\cite{Dymnikova:1992ux,Easson:2002tg,Hayward:2005gi,Shankaranarayanan:2003qm,Berej:2006cc,Bronnikov:2005gm, Dymnikova:2015yma,Calza:2024fzo,Davies:2024ysj,Asmanoglu:2025agc}. In these models, the spacetime geometry smoothly transitions from an asymptotically Schwarzschild or Reissner-Nordström solution at large radii to a de Sitter geometry as $r \to 0$. While conceptually appealing, a significant drawback of these solutions is their reliance on \emph{ad hoc} matter field configurations. The Bardeen BH, for instance, can be interpreted as an exact solution to Einstein's equations coupled to a specific form of non-linear electrodynamics~\cite{Ayon-Beato:1998hmi}. In general, such regular solutions require the violation of the Strong Energy Condition, often demanding matter sources with exotic properties that are not derived from a fundamental theory~\cite{Dymnikova:1992ux,Easson:2002tg,Hayward:2005gi,Shankaranarayanan:2003qm,Berej:2006cc,Calza:2024fzo,Davies:2024ysj,Asmanoglu:2025agc}.

This leads to a crucial question: Is the requirement of exotic matter an unavoidable neccessity for singularity resolution, or does the issue lie deeper? 
We argue that the fundamental limitation lies in two related aspects of standard matter formulations. First, 
the standard formulation of matter itself, where actions are built from ``canonical" Lagrangians that depend \emph{linearly} on the kinetic terms. In other words, the assumption of \emph{linear response} is inherent to canonical field theories. This is a direct analogue of \emph{Hooke's law in elasticity} --- an excellent approximation for small strains but destined to fail under extreme stress. The gravitational collapse into a singularity represents the ultimate \emph{large strain} regime for field theory, where the linear approximation must be superseded by a non-linear UV-completion that can capture the fundamental physics of high gradients~\cite{Falkowski:2023hsg}.

From this perspective, the central singularity is a pathology of pushing an ineffective, low-energy theory --- the canonical kinetic term --- beyond its regime of validity. To show this explicitly, we consider a scalar field $\phi$. In the framework of Effective Field Theory (EFT)~\cite{Weinberg:1978kz,Georgi:1993mps,Burgess:2007pt}, the action for a scalar field is not simply \(\mathcal{L}_\phi = -X \equiv \partial^\mu\phi \partial_\mu \phi/2\) but is expected to be an infinite series of higher-dimensional operators~\cite{Penco:2020kvy,Shankaranarayanan:2022wbx,Mandal:2025xuc}:
\[
\mathcal{L}_\phi^{\text{EFT}} = -X + \frac{\alpha_2}{\Lambda^4}X^2 + \frac{\alpha_3}{\Lambda^8}X^3 + \dots,
\]
where \(\Lambda\) is the UV cutoff scale and $\alpha_2, \alpha_3$ are dimensionless coupling constants. (We work with the metric signature $(-,+,+,+)$ and set $c = \hbar = 1$.)
The canonical kinetic term is merely the leading-order operator, valid only when field gradients are small, 
\(X/\Lambda^4 \ll 1\). However, in the high-curvature environment of a BH's core, gravitational collapse drives field gradients to extreme values, \(X \to \infty\)~\cite{Mandal:2025xuc}. In this regime, the EFT expansion breaks down catastrophically; the infinite series of higher-order terms becomes dominant, and truncating the series at any finite order is unjustified~\cite{Shankaranarayanan:2022wbx,Mandal:2025xuc}. The resulting physics is not only unreliable but often unphysical, leading to pathologies like ghost and gradient instabilities~\cite{Barth:1983hb,Simon:1990ic}.

The central singularity of a BH can therefore be viewed as a direct consequence of extrapolating a low-energy effective theory (the canonical scalar) into a regime where its basic assumptions are violated. To understand the fate of gravitational collapse, one must specify the UV-completion of the scalar field theory. A guiding principle for such modifications comes from Born-Infeld non-linear electrodynamics, which was introduced to resolve the infinite self-energy of the point electron~\cite{Born:1934gh,Dirac:1962iy}. This concept was generalized to scalar fields in cosmology (k-inflation)~\cite{Armendariz-Picon:1999hyi} and string theory~\cite{Callan:1997kz}, leading to the Dirac-Born-Infeld (DBI) action. 
In other words, the DBI action represents a specific, and highly constrained non-linear field theory valid until the energy scale $\Lambda$~\cite{Kluson:2000iy,Gibbons:2000hf,Sen:2002an}:
\begin{equation}
\mathcal{L}_\phi = V(\phi)\left[1-\sqrt{1+\frac{2X}{\Lambda^4}}\right],~X = -\frac{1}{2}g^{\mu\nu}\nabla_\mu\phi\nabla_\nu\phi.
\label{eq:DBIaction}
\end{equation}
Specifically, the above action is not just the first few terms of the EFT expansion; it is a \emph{resummation} of the entire infinite series to all orders in \(X\) 
valid until the energy scale $\Lambda$. Its structure is uniquely fixed by underlying symmetries, making it a robust and predictive framework for exploring the high-gradient regime. 
Its defining feature is a \emph{speed limit} for field gradients, \(|(\nabla\phi)^2| < \Lambda^4\), which naturally suppresses the large gradients 
(until the energy scale $\Lambda$) that would otherwise lead to a singularity. Such non-linear k-essence theories are now central to modern physics, from inflationary models to effective field theories of massive gravity (e.g., Galileons)~\cite{deRham:2010kj,Silverstein:2003hf,Alishahiha:2004eh,Miranda:2012rm}. 

We therefore take the logical step of applying this well-motivated UV-completion to the strong-gravity environment of a BH interior, where controlling gradients is paramount. 
However, our analysis reveals a second, more subtle limitation: not all non-linear completions are equally effective. We demonstrate that while the full Dirac-Born-Infeld action --- which resums the entire infinite series of higher-order kinetic terms --- can indeed support a non-singular black hole, this is only possible for its phantom branch. This result provides a precise characterization of the matter sector required for singularity resolution: it must not only incorporate non-linear kinetic terms, but those of the phantom DBI type.

While fundamental phantom fields are problematic, their appearance as an \emph{effective description} is well-established in modern field theory~\cite{Callan:1969sn,Arkani-Hamed:2003pdi}. For instance, non-canonical (k-essence) actions~\cite{Armendariz-Picon:1999hyi} with phantom-like properties emerge from higher-dimensional theories, non-linear sigma models, or from integrating out auxiliary fields in a fundamental theory~\cite{Callan:1969sn,Arkani-Hamed:2003pdi} (see  Appendix~\ref{phantom_DBI} for details). This suggests our solution represents the stable, classical endpoint of a more fundamental, non-pathological parent theory, rather than being sourced by a fundamental phantom. The resulting spacetime is asymptotically flat, evades the no-hair theorems by possessing non-trivial scalar hair, and features a regular 2-sphere of finite area at its core --- providing a plausible resolution of the central singularity.

\section{Field Equations}
We consider the 4-D action where $\phi$ is minimally coupled to gravity:
\begin{equation}
\label{DBI:Lagrangian}
S=\int d^4x\sqrt{-g}\left[{R}/{2\kappa^2}+\epsilon \, 
\mathcal{L}_\phi \right] \, ,
\end{equation}
where $\kappa^2=8\pi G$,  and $G$ is the Newton's constant. The low-energy limit $X \ll \Lambda^4$ gives $
\mathcal{L}_\phi \approx -\epsilon X {V(\phi)}/{\Lambda^{4}}$. Thus, $\epsilon = -1$ 
(and $V(\phi)=\Lambda^{4}$) corresponds to the standard (non-phantom) scalar field, while $\epsilon = 1$ represents a phantom field. We keep $\epsilon$ explicit to determine which case supports the solution. 
%
Varying the above action w.r.t $\phi$ and $g_{\mu\nu}$ yields:
\begin{eqnarray}
\label{Phi:EOM}
\nabla_{\mu}\left(\frac{V(\phi)\nabla^{\mu}\phi}{\sqrt{1+\frac{2X}{\Lambda^4}}}\right) -  
\Lambda^{4}V^{\prime}(\phi)\left[1-\sqrt{1+\frac{2X}{\Lambda^4}}\right] & = & 0,~~\\
\label{DBI:metric:Eom}
G^{\mu}_{\nu} - 
{\kappa^2} \, \epsilon
\left[ \mathcal{L}_\phi \, 
\delta^{\mu}_{\nu} 
- {\frac{V(\phi)\left(\nabla^{\mu}\phi\right)(\nabla_{\nu}\phi)}{\Lambda^4 \sqrt{1+\frac{2X}{\Lambda^4}}}}
\right] & = & 0.
\end{eqnarray}
%
where prime denotes derivative w.r.t $\phi$.

We seek a static, spherically symmetric solution using the ansatz:
%
\begin{equation}
\label{metric:ansatz}
ds^2=-f(r)dt^2+ {dr^2}/{f(r)}+\rho^2(r) \, d\Omega^2 \,.
\end{equation}
where $f(r), \rho(r)$ are continuous, differentiable
functions of $r$~\footnote{$\rho$ is the areal radius and is a scalar quantity that depends on the coordinates of the normal space-time}, and $d\Omega^2$ is the metric on $S^2$.
For this metric, $T^{t}_{t} = T^{\theta}_{\theta}$, which implies $G^{t}_{t} - G^{\theta}_{\theta} = 0$. The Bianchi identity ($\nabla^\mu G_{\mu}^{\nu} = 0$) ensures the scalar EOM \eqref{Phi:EOM} is redundant. 
This leaves three independent equations for four unknown functions ($f, \rho, \phi, V$), rendering the system underdetermined. This freedom allows us to set forth a physically-motivated ansatz. \emph{For both cases $\epsilon = \pm 1$}, we seek a solution that is asymptotically flat ($\rho(r) \to r$ and $f(r) \to 1 - 2GM/r$ as $r \to \infty$) and regular at the center.

\section{Solution and Regularity}
For the metric ansatz \eqref{metric:ansatz}, the scalar EOM \eqref{Phi:EOM} and Einstein equations $G^{t}_{t} - G^{\theta}_{\theta} = 0$ are:
\begin{eqnarray}
\label{Phi:EOM2}
\frac{1}{\rho^2} \frac{d}{dr}\bigg[\frac{\rho^{2}f\phi^{\prime}V(\phi)}{\sqrt{1-\frac{f(\phi^{\prime})^2}{\Lambda^4}}}\bigg] &=& 
\Lambda^4V^{\prime}(\phi)\left[1-\sqrt{1-\frac{f(r)(\phi^{\prime})^2}{\Lambda^4}}\right]~~~ \\
\label{eq:EFE2}
\frac{\rho^{2}(r)f^{\prime\prime}}{2} &=& 
f\rho\rho^{\prime\prime}-1+f(r)(\rho^{\prime}(r))^2 \, .
\end{eqnarray}
Crucially, Eq.~\eqref{eq:EFE2} is independent of $\phi$ and $\epsilon$,  allowing us to solve the geometry first.
We begin with the ansatz $\rho(r)=(r^{p}+a^{p})^{\frac{1}{p}}$ and infer that for an asymptotically flat metric solution, the only possible choice is $p=2$. (Details in Appendix~\ref{exactsol}.)
Hence, we consider the ansatz $\rho(r)=\sqrt{r^2+a^2}$, which is regular and defines a minimal 2-sphere of area $4\pi a^2$ at $r=0$.  Eq.~\eqref{eq:EFE2} then simplifies to a linear ODE, $(r^2+a^2)f^{\prime\prime}-2f+2=0$. The unique solution that is asymptotically flat ($f \to 1 - 2GM/r$ as $r \to \infty$) is:
\begin{equation}
\label{eq:f(r)_gen}
    f(r)=1+C_{1} \rho^2(r)+\frac{C_{2}}{2a^2}\left[r - \frac{\rho^2(r)}{a}\tan^{-1}\left(\frac{a}{r}\right)\right].
\end{equation}
where $C_1, C_2$ are integration constants. We impose asymptotic flatness ($f \to 1$ as $r \to \infty$) and match the ADM mass $M$ ($f(r) \approx 1 - 2GM/r$). This fixes $C_1 = 0$ and $C_2 = 6GM$, yielding our first key result:
\begin{eqnarray}
\label{final:f(r)}
 f(r)=1 + \frac{3GM}{a} \left[ 
 \frac{r}{a} - \frac{\rho^2(r)}{a^2}   
\tan^{-1}\left(\frac{a}{r}\right) \right].~
\end{eqnarray}
It is crucial to notice that irrespective of the nature of the DBI field (regular or phantom), this metric is regular everywhere. We establish geodesic completeness by verifying: (i) metric functions are smooth and positive-definite for all \(r \in [0,\infty)\); (ii) all curvature invariants remain finite; (iii) ensuring radial null geodesics reach the center at finite affine parameter and are smoothly extendable. (Details in Appendix~\ref{Invariants_Completeness} and~\cite{DBI-Zenodo}.)

Given this regular geometry, we must determine the form of $\phi(r)$ using the remaining EOMs.
%
%
In order to solve for $\phi(r)$, we first define an auxiliary function:
\begin{equation}
\label{def:Q}
Q(r) = \sqrt{1 + \frac{2X}{\Lambda^4}} = \sqrt{1 - \frac{f(r)(\phi')^2}{\Lambda^4}}~.
\end{equation}
The metric field equations in terms of the auxiliary function $Q(r)$ have the following form:
\begin{align}
    G^{t}_{t}&=\epsilon\kappa^2V(\phi)\left(1-Q(r)\right) \\
    \label{Gtt_minus_Grr}
    G^{t}_{t}-G^{r}_{r}&=\epsilon\kappa^{2}V(\phi)\left(\frac{1}{Q(r)}-Q(r)\right)~.
\end{align}

\noindent Taking the ratio of the above expressions, we obtain a simple expression for $Q(r)$: $Q(r)=-G^{t}_{t}/{G^{r}_{r}}$.
%
%
Using \eqref{def:Q} and the final expression of $Q(r)$, we obtain the following integral representation of $\phi(r)$:
\begin{align}
    \label{IntExp:phi}
    \phi(r)= \pm \int dr \sqrt{\frac{\Lambda^4}{f(r)}\left(\frac{(G^{r}_{r})^2-(G^{t}_{t})^2}{(G^{r}_{r})^2}\right)}~~,
\end{align}
%
The above expression in \eqref{IntExp:phi} is our second key result.


\color{black}
\section{Choice of $\epsilon$}
We now have an exact, regular solution, but we must check its physical consistency. 
Eq. \eqref{Gtt_minus_Grr} provides the crucial test.
The left-hand side (LHS) depends only on our geometric ansatz:
\begin{equation}
\text{LHS} = \frac{2 \rho^{\prime\prime}}{\rho} = \frac{2 a^2}{(r^2+a^2)^2} > 0 \, .
\end{equation}
The right-hand side (RHS) depends on the scalar field properties:
\begin{equation}
\text{RHS} = \frac{\epsilon\kappa^2V(\phi(r))(\phi^{\prime}(r))^2}{\Lambda^4 Q(r)}~.
\end{equation}
The ratio $V(\phi(r))/Q(r)$ is strictly positive and $(\phi^{\prime}(r))^2$ is also positive.
Hence, $\text{sign(RHS)} = \text{sign}(\epsilon)$. For the equality to hold, we must have:
\begin{equation}
    \epsilon = +1
\end{equation}
This is a powerful result. It demonstrates that our non-singular solution can only be supported by the \emph{phantom DBI action} ($\epsilon = +1$). The standard (non-phantom) case ($\epsilon = -1$) is inconsistent with this geometry.


The apparent phantom nature of the scalar field in our model should be understood within the framework of Effective Field Theory (EFT), where such non-canonical Lagrangians emerge as low-energy descriptions of more fundamental multi-field theories. Specifically, the DBI action structure can be derived from a Non-Linear Sigma Model (NLSM) describing a dynamical field $\phi$ coupled to an auxiliary field $\chi$ via a Lagrangian of the form $\mathcal{L} \sim (1 + X/\Lambda^4)\chi^a + \chi^{b}$~\cite{Callan:1969sn, Weinberg:1978kz, Armendariz-Picon:1999hyi}. Since the auxiliary field lacks a kinetic term, integrating it out via its algebraic equation of motion generates a non-trivial effective action for the dynamical field, $\mathcal{L}_{\text{eff}} \propto \sqrt{1 + X/\Lambda^4}$ ((see Appendix~\ref{phantom_DBI} for details). When coupled to gravity as presented in our action, the low-energy expansion ($X \ll \Lambda^4$) of this effective theory yields a kinetic term with a reversed sign, naturally giving rise to the phantom behavior. Thus, the phantom nature is not necessarily a fundamental property but rather an emergent feature of the effective description valid below the cutoff scale $\Lambda$.

\section{Horizon Structure and Constraints}
\begin{figure}[!htb] 
\begin{center}
\includegraphics[width=0.65\linewidth]{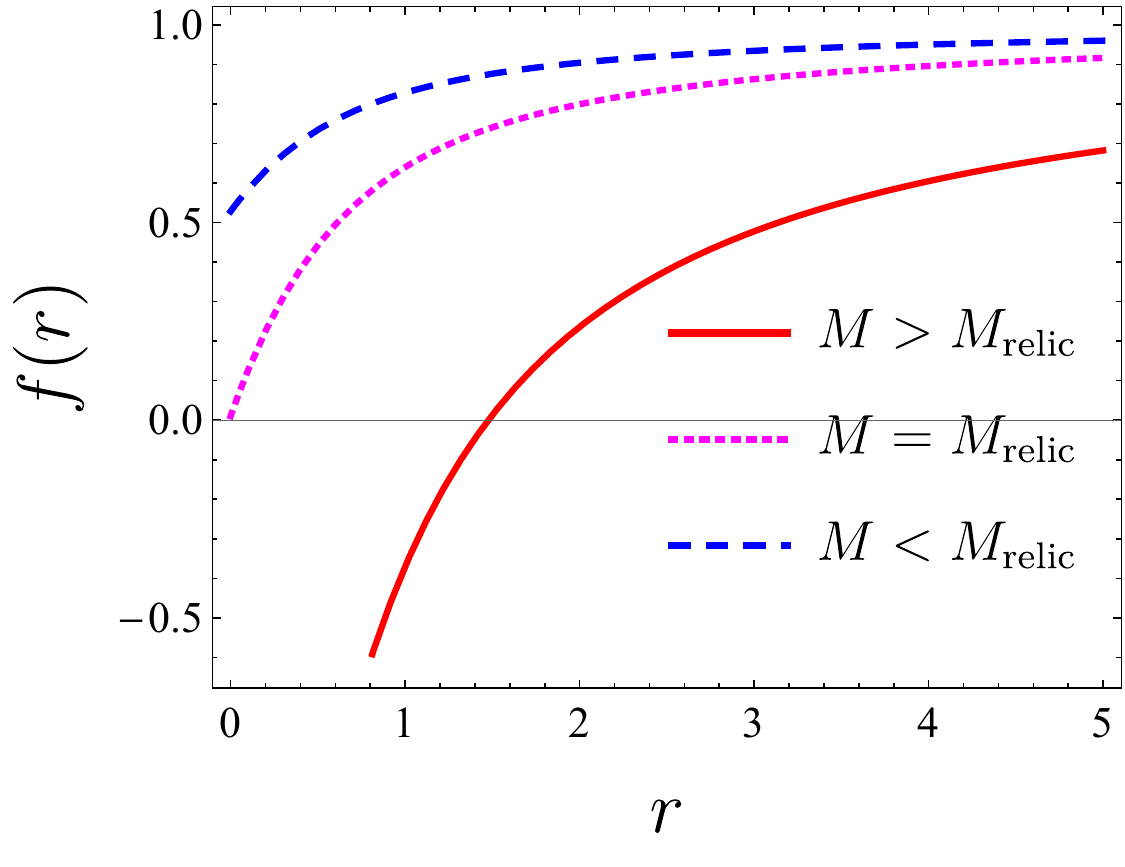}
\caption{The metric function $f(r)$ for the three mass regimes (where $a = G = 1$), showing the non-extremal BH (red), the stable extremal relic (Magenta), and the horizon-less regular object (blue) .}
\label{fig:fr}
\end{center}
\end{figure}

The solution \eqref{final:f(r)} represents a BH only if $f(r)=0$ has positive roots, which define the event horizon(s). The horizon structure depends on the sign of $f(r)$ at the core $r=0$. Using $\lim_{r\to 0} \tan^{-1}(a/r) = \pi/2$, we find $f(0) = 1 - \frac{3GM\pi}{2a}$. 
As shown in Fig.~\ref{fig:fr}, we have three distinct mass regimes:
\begin{enumerate}
\item \textbf{Non-extremal BH ($M > {2a}/{(3\pi G)}$):} Here, $f(0) < 0$. As $f(r)$ must cross from negative to positive, there is at least one horizon $r_h > 0$. This is a regular BH.
\item \textbf{Extremal BH ($M_{\rm relic} = {2a}/{3\pi G}$):} For $f(0) = 0$, the horizon coincides with the regular core ($r_h = 0$). This defines the minimum relic mass $M_{\rm relic} = {2a}/{3G\pi}$.

\item \textbf{Regular Object ($M < {2a}/{(3\pi G)}$):}  
In this case, $f(0)$ is positive\footnote{There is also a case where the function starts at a local maximum at $r=0$ and may dip to form two or multiple horizons, depending on the mass-to-scale ratio. We do not consider this case here.}. $f(r)$ starts positive at $r=0$, and $f(\infty)$ is also positive. For a low enough mass, it's possible for $f(r)$ to be positive everywhere. In this range, there is \emph{no event horizon}. This is a horizon-less 
regular, particle-like object~\cite{Simpson:2018tsi}. 

This horizon-less object can be considered \emph{wormhole-like}~\cite{Morris:1988cz,Visser:2002ua,Shoshany:2019yoz}.  The solution for $r \in [0, \infty)$ can be interpreted as representing one ``side" of a traversable wormhole. One could analytically continue the solution to $r \in (-\infty, \infty)$, and the throat at $r=0$ would then connect two separate asymptotically flat spacetimes. 
In the rest of the analysis, we will not consider these objects and will focus on BH and their properties.

\end{enumerate}

%
We can derive a physical relation between the regularization parameter $a$ and the DBI scale $\Lambda$ by analyzing the energy density of the scalar field at the core of the regular black hole. Using the Einstein field equations, the energy density $\rho_E = -T^t_t \sim G^t_t$ at the center ($r=0$) is found to be finite (see Appendix~\ref{relation_between_parameters} for details):
\begin{equation}\rho_{\text{core}} \sim \frac{3\pi GM}{\kappa^2 a^3} = \frac{3M}{8 a^3}
\end{equation}
where we have used $\kappa^2 = 8\pi G$. In DBI theories, the scale $\Lambda^4$ represents characteristic energy density of the field. Identifying the maximum core density with this fundamental scale ($\rho_{\text{core}} \sim \Lambda^4$) yields the relation:
\begin{equation}
\label{eq:a-LambdaRel}
a \simeq \left( {3M}/{(8\Lambda^4)} \right)^{1/3}\end{equation}
This indicates that $a$ is determined by the interplay between the total mass $M$ and the maximum field density $\Lambda^4$ allowed by the theory. For a fixed scale $\Lambda$, the core size $a$ scales with the cube root of the mass.
  
\section{BH Thermodynamics and Implications for PBH}
Let us now look at the implications for the BH thermodynamics. Consider the case where $2 GM \gg a$ where we can expand $f(r)$~\eqref{final:f(r)} in powers of $a$. The metric function \eqref{final:f(r)} becomes:
\begin{equation}
f(r) \approx 1 - \frac{2GM}{r} + \frac{2GMa^2}{5r^3} + \mathcal{O}(a^4)
\end{equation}
This is the Schwarzschild metric plus a small $a^2$ correction. Solving $f(r_h)=0$ for the horizon radius gives $r_h \approx 2GM - a^2/(10GM)$. The surface gravity is $\kappa_s = \frac{1}{2}f'(r_h)$, leading to a Hawking temperature~\cite{Hawking:1975vcx}:
\begin{equation}
T_H = {\kappa_s}/{(2\pi k_B)} \approx T_S \left(1 - {a^2}/{(20(GM)^2)} \right)
\end{equation}
where $T_S = 1/(8\pi GM k_B)$ is the Hawking temperature of a Schwarzschild BH and $k_B$ is the Boltzmann constant. The BH area is $A = 4\pi\rho(r_h)^2 = 4\pi(r_h^2+a^2) \approx A_S ( 1 + \frac{3a^2}{20(GM)^2} )$, where $A_S = 4\pi (2GM)^2$.
The evaporation rate (Power) $P = -dM/dt \propto A T_H^4$, is then~\cite{Page:1976df}:
\begin{equation}
P \approx P_S \left( 1 - {a^2}/{(20(GM)^2)} \right)
\end{equation}
where $P_S$ is the evaporation rate for a Schwarzschild BH. 
This result has two crucial implications for Primordial BHs (PBHs), one perturbative and one non-perturbative. The perturbative analysis explicitly shows that the evaporation rate is slightly slower ($P < P_S$), hence, the BH's lifetime $\tau \propto M^3/P$ is slightly \emph{longer} than a Schwarzschild PBH. This marginally shifts the mass constraints. For a similar result, see Refs.~\cite{Davies:2024ysj,Asmanoglu:2025agc}. 

However, an interesting implication is the \emph{non-perturbative regime} when $M$ is no longer large and the above expansion in $a$ is not valid. Assuming $\Lambda \sim 10^{17} GeV$ all through and using the relation \eqref{eq:a-LambdaRel}, the mass $M$ will shrink until it reaches the \emph{extremal mass}:
%
\begin{equation}
M_{\rm relic} \sim 8 \times M_{\rm Planck}^3/\Lambda^2 \sim 1~{\rm g} \, . 
\label{eq:RelicMass}
\end{equation}
whose horizon radius is less than the physical radius $a$. It is important to note that this analysis assumes the validity of this theory till the end stages of evaporation. To verify the validity, one needs to do a non-perturbative analysis which is beyond the scope of this work.

Given the above caveat, our regular BH solution has important implications for PBH dark matter~\cite{Carr:2020xqk}. Evaporating Schwarzschild BHs with initial masses $M < 10^{15}\,$g are ruled out as dark matter because they would have evaporated by today, producing an observable gamma-ray background. Our regular DBI BHs completely evade this constraint. PBHs with arbitrarily small initial masses $(M > M_{\rm relic}$) could have formed, evaporated down to a remnant weighing $\sim 1~{\rm g}$, and now populate the Universe as the cold dark matter\footnote{It is possible that the regular object that were formed at the early Universe can also contribute as dark matter which is not considered in this analysis.}. For a similar result, see Refs.~\cite{Davies:2024ysj,Asmanoglu:2025agc}. 
\section{Evasion of no-hair theorem}
The derivative of the scalar field 
vanishes at infinity 
 and hence the scalar``hair" does not contribute to the stress-tensor at infinity, consistent with asymptotic flatness. This result evades the classical no-hair theorems~\cite{Bekenstein:1972ny,Hawking:1972qk,Bekenstein:1998aw,Sotiriou:2011dz,Herdeiro:2015waa}, which are known to fail for non-linear matter or for theories with higher-order corrections, such as Einstein-scalar-Gauss-Bonnet theories which permit a variety of \emph{hairy BHs}~\cite{Antoniou:2017acq, Doneva:2017bvd, Silva:2017uqg}. 
 According to the classification of scalar hair provided in Ref.~\cite{Bakopoulos:2023fmv}, a \emph{primary
hair} implies a new, independent quantum number (scalar charge) that is not fixed by the other conserved charges~\cite{Hui:2012qt,Capuano:2023yyh}. Hence, the solution here corresponds to secondary hair.

The mechanism for both singularity resolution and no-hair evasion is intrinsic to the non-linear kinetic structure of the phantom DBI field. The Lagrangian \eqref{eq:DBIaction} imposes a fundamental \emph{speed limit} on field gradients, set by the scale $\Lambda$. As gravitational collapse drives the scalar field toward a would-be singularity, its gradient approaches this maximum value. The field responds by dynamically resisting further compression. 
The regularization capability of the phantom DBI action \eqref{eq:DBIaction} stems from its intrinsic \emph{kinetic speed limit}. The requirement that the argument of the square root remains positive imposes a fundamental bound~\cite{Armendariz-Picon:1999hyi}:
\[
1 + {2X}/{\Lambda^4} > 0 \quad \Longrightarrow \quad g^{\mu\nu}\partial_\mu\phi\partial_\nu\phi < \Lambda^4.
\]
This ensures the field's gradient cannot exceed the EFT scale \(\Lambda\). During gravitational collapse, as the field is compressed and its gradient approaches this maximum, the factor \(\gamma \equiv (1 + 2X/\Lambda^4)^{-1/2}\) grows large. This is directly analogous to the Lorentz factor in special relativity, where a particle's energy diverges as its speed approaches the speed of light, making further acceleration impossible. Similarly, the DBI scalar's effective pressure~\cite{Armendariz-Picon:1999hyi}:
\[
P_{\rm eff} \sim \Lambda^4 (\gamma^2 + \gamma - 2)/{\gamma},
\]
diverges as \(\gamma \to \infty\), generating a powerful, dynamically emergent repulsive force.
Physically, the phantom DBI field develops enormous effective pressure when its kinetic terms become dominant, halting gravitational collapse at a finite radius and supporting a stable, non-singular core. It is important to note that this is not an externally imposed cosmological constant core but a \emph{dynamical} response intrinsic to the field's microphysics. 

The presence of non-trivial scalar hair will also imprint unique signatures on gravitational waves. Unlike vacuum GR BHs, which only admit quadrupole radiation, the scalar field introduces new channels for energy loss, potentially including dipole radiation 
(scaling as $v^{-2}$ relative to quadrupole)~\cite{Gonzalez:2025yjm}. This additional energy loss channel accelerates the inspiral phase, potentially detectable by LISA or Einstein Telescope as a phase shift in the waveform. This modifies the BH's tidal deformability (Love numbers) and its quasinormal mode spectrum (see Appendix), providing distinct, testable signatures in the inspiral and post-merger ringdown, respectively~\cite{Berti:2015itd,Cardoso:2019rvt} 
(see details in Appendix~\ref{QNMs}). These deviations from the Kerr paradigm are primary targets for next-generation observatories like Cosmic Explorer (CE)~\cite{Evans:2021gyd,Hall:2022dik} and the Einstein Telescope (ET)~\cite{Branchesi:2023mws,ET:2025xjr}, which are designed to probe such modifications to the compact object's nature.

\section{Conclusions and Implications}
The discovery of a stable, regular BH sourced by an effective phantom DBI field has important implications across multiple frontiers of physics.
First, our analysis demonstrates that \emph{kinetic regularization} ---where non-linear field self-interactions generate a repulsive force at high gradients --- provides a robust, well-motivated principle for singularity resolution.  This addresses the fundamental limitation of canonical field theories, which fail catastrophically in the large-strain regime of gravitational collapse. Furthermore, we have shown that singularity resolution is not guaranteed by an arbitrary non-linear completion; \emph{it necessarily requires the specific structure of the phantom DBI action} $(\epsilon = 1)$. Unlike models requiring \emph{ad hoc} exotic matter~\cite{Ayon-Beato:1998hmi} or speculative quantum gravity~\cite{Bojowald:2024lhr}, 
our mechanism emerges naturally from the EFT of a scalar field with maximum energy $\Lambda$.

Second, this solution provides a natural and robust candidate for \emph{PBH dark matter}~\cite{Carr:2020xqk}. The existence of a remnant weighing 
(\(M_{\text{relic}} \sim 1 \text{g}\)) is a direct consequence of the phantom DBI dynamics. This evades all standard gamma-ray constraints on evaporating PBHs, resurrecting a vast, previously forbidden mass window for light PBHs to constitute the cold dark matter of the Universe.  

Finally, the presence of non-trivial scalar hair makes this model phenomenologically rich and testable. The scalar field introduces new channels for energy loss (e.g., dipole radiation) and modifies tidal deformabilities and quasi-normal modes, providing distinct gravitational-wave signatures. These deviations from the Kerr paradigm are primary targets for next-generation observatories like Cosmic Explorer and the Einstein Telescope~\cite{Evans:2021gyd,Hall:2022dik,Branchesi:2023mws,ET:2025xjr}. The solution also offers a framework for interpreting high-mass mergers in the upper mass gap ~\cite{LIGOScientific:2018mvr,KAGRA:2021vkt} as potential PBH mergers.

This work establishes non-linear, kinetic-limited field theories—already proven in cosmology—as an essential new pathway for understanding the ultimate fate of gravitational collapse. The necessity of the phantom branch points to a deeper interplay between UV-completions of matter and the resolution of spacetime singularities.

An important future direction is the generalization of this exact solution to the rotating regular BHs. Constructing rotating regular counterparts is notoriously difficult, as naive applications of the Newman-Janis algorithm to non-vacuum solutions often lead to pathologies or require complex matter sources~\cite{Erbin:2014aja,BenAchour:2025lkx,Fernandes:2026rjs}. A rigorous extension of our DBI framework to axisymmetry would be crucial for generating precise waveform templates for future GW observatories.

While our thermodynamic analysis is restricted to the $GM \gg a$ limit, determining the precise cosmological abundance and observational constraints requires a full numerical analysis of the evaporation history. A detailed study, analogous to those performed with public codes like \emph{BlackHawk}~\cite{Arbey:2019mbc,Auffinger:2022khh}, is currently under investigation and will be presented elsewhere.

\noindent\emph{\underline{Acknowledgements:}} 
We dedicate this work to the memory of Prof. Naresh Dadhich, who, even in his thoughtful critiques of phenomenological regular black hole models, pushed us toward greater mathematical and physical clarity. We honor his contributions and mourn his sudden loss during a research visit to Beijing. The authors are grateful to I. Chakraborty, S. M. Chandran, A. Chowdhury, P. G. Christopher,  K. Hari, N. Jaiswal, A. Kushwaha, S. Jana, S. Mandal, K. Rajeev, and S. Xavier for their valuable discussions and feedback on the earlier draft. The work is supported by ANRF-Advanced Research Grant (ANRF/ARG/2025/001514/PS).  The MHRD fellowship at IIT Bombay financially supports TP.
\appendix

\section{DIRAC-BORN-INFELD (DBI) ACTION}
\label{DBI_Action_Appendix}
For reproducibility, the full Mathematica notebook (.nb) used to derive and verify all calculations in this work is permanently archived at \href{https://zenodo.org/records/17659601}{Zenodo}.

We consider the 4-D action where the phantom DBI scalar field $(\epsilon = 1)$, $\phi$ is minimally coupled to gravity: 
\begin{equation}
    \label{DBI:Lagrangian}
    S=\int d^4x\sqrt{-g}\bigg[\frac{R}{2\kappa^2}+
    V(\phi)\bigg(1-\sqrt{1+\frac{2X}{\Lambda^4}}\bigg)\bigg]~, \quad
    X=-\frac{1}{2}g^{\mu\nu}\nabla_{\mu}\phi\nabla_{\nu}\phi~.
\end{equation}
where $\kappa^2=8\pi G$, $G$ is the Newton's constant, $\Lambda$ is the intrinsic energy scale in the theory, and $X$ is the kinetic energy of the DBI scalar field. We work with the metric signature $(-,+,+,+)$ and set $c=\hbar=1$. 
Varying the above action w.r.t $\phi$ yields:
\begin{equation}
     \label{SM:Phi:EOM}
\nabla_{\mu}\left(\frac{V(\phi)\nabla^{\mu}\phi}{\sqrt{1+\frac{2X}{\Lambda^4}}}\right)=V^{\prime}(\phi)\Lambda^4\left(1-\sqrt{1+\frac{2X}{\Lambda^4}}\right)~.
\end{equation}
Varying the action \eqref{DBI:Lagrangian} w.r.t the metric $g_{\mu\nu}$, we obtain the following EOM:
\begin{equation}
    \label{SM:DBI:metric:Eom}
    \frac{G_{\mu\nu}}{\kappa^2}=V(\phi)\left(1-\sqrt{1+\frac{2X}{\Lambda^4}}\right)g_{\mu\nu}-\frac{V(\phi)\left(\nabla_{\mu}\phi\right)(\nabla_{\nu}\phi)}{\Lambda^{4}\sqrt{1+\frac{2X}{\Lambda^4}}}~.
\end{equation}
The static, spherically symmetric metric in 4-D can be written in the following form:
\begin{equation}
    \label{SM:metric:ansatz}
    ds^2=-f(r)dt^2+dr^2/f(r)+\rho^{2}(r)(d\theta^2+\sin^2\theta d\phi^2)~,
\end{equation} 
where $f(r)$ and $\rho(r)$ are unknown (continuous, differentiable) functions of the radial coordinate $r$. Substituting the above line-element in the Einstein's equations \eqref{SM:DBI:metric:Eom} we see that $T^{t}_{t}=T^{\theta}_{\theta}=T^{\phi}_{\phi}$ leading to $G^{t}_{t}-G^{\theta}_{\theta}=0$. Crucially, the Bianchi identity $(\nabla^{\mu}G^{\nu}_{\mu}=0)$ ensures that the DBI scalar field equation \eqref{SM:Phi:EOM} is not independent of the metric equations \eqref{SM:DBI:metric:Eom}. 
This leaves us with only three independent equations of motion but four unknown functions: $f(r)$, $\rho(r)$, and $\phi(r)$, $V(\phi(r))$. The system is thus \emph{underdetermined}, allowing us to specify one function as a physically motivated ansatz. We look for a solution that is asymptotically flat $(\rho(r)\rightarrow r)$ and $(f(r)\rightarrow 1-2GM/r$ as $r\rightarrow \infty$) and regular at the center.

\section{Exact solution of the field equations}
\label{exactsol}

\subsection{Solving for $f(r)$}
For the metric ansatz \eqref{SM:metric:ansatz}, the Einstein equation $G^{t}_{t}-G^{\theta}_{\theta}=0$ takes the following form
\begin{equation}
\label{SM:Einstein:eq1}
\frac{1}{\rho^{2}(r)}\bigg[\frac{\rho^{2}(r)f^{\prime\prime}}{2}-f\rho\rho^{\prime\prime}+1-f(r)(\rho^{\prime}(r))^2\bigg]=0~.
\end{equation}
The absence of the scalar field $\phi$ in the above equation allows us to solve the geometry first. 
We first make the generic ansatz $\rho(r)=(r^{p}+a^{p})^{\frac{1}{p}}$, where $p$ is any real number. With the choice of the generic ansatz for $\rho(r)$, we obtain the following expressions for derivatives of $\rho(r)$: 
\begin{eqnarray}
    \rho^{\prime}(r)=r^{p-1}(r^{p}+a^{p})^{\frac{1}{p}-1}\hspace{0.1cm};\hspace{0.1cm}\rho^{\prime\prime}(r)=a^{p}(p-1)r^{p-2}(r^{p}+a^{p})^{\frac{1}{p}-2}
\end{eqnarray}


Substituting the above expressions in \eqref{SM:Einstein:eq1}, we obtain the following ODE:
\begin{eqnarray}
    \frac{(r^{p}+a^{p})^{\frac{2}{p}}f^{\prime\prime}(r)}{2}+1-f(r)r^{p-2}(r^{p}+a^{p})^{\frac{2}{p}-2}\left[r^{p}+a^{p}(p-1)\right]=0
\end{eqnarray}

\noindent From the above ODE, we infer that $f(r)=1$ is only a solution when $p =2$. Hence, in order to obtain a consistent asymptotically flat solution, we need to seet $p = 2$, which defines a minimal 2-sphere at $r=0$ with area $4\pi a^2$. Eq. \eqref{SM:Einstein:eq1} simplifies to the following linear ODE with the ansatz for $\rho$:
\begin{equation}
(r^2+a^2)f^{\prime\prime}(r)-2f(r)+2=0~.
\end{equation} The general solution to the above ODE is:
\begin{equation}
f(r)=1+C_{1}(r^2+a^2)+\frac{C_{2}}{2a}\bigg(\frac{r}{a}-\frac{(r^2+a^2)}{a^2}\tan^{-1}\left(\frac{a}{r}\right)\bigg)~,
\end{equation}
where $C_{1}$ and $C_{2}$ are integration constants.
We impose asymptotic flatness and match the ADM mass M $(f(r)\approx 1-2GM/r$ as $r\rightarrow \infty$). In the $r\rightarrow \infty$ limit we find
\begin{equation}
f(r)\simeq 1+ C_{1} (r^2+a^2)-\frac{C_{2}}{3r}+\mathcal{O}\left(\frac{1}{r^3}\right)~.
\end{equation}
Hence from the requirement of asymptotic flatness and matching the ADM mass, we obtain $C_{1}=0\hspace{0.1cm},\hspace{0.1cm}C_{2}=6GM$. The resulting expression for the metric function $f(r)$ is 
\begin{equation}
\label{SM:f(r):exp}
f(r)=1+ \frac{3GM}{a}\bigg[\frac{r}{a}-\frac{r^2+a^2}{a^2}\tan^{-1}\left(\frac{a}{r}\right)\bigg]~.
\end{equation}

\subsection{Solving for $\phi(r)$}
In this subsection, we solve for the DBI scalar field. For a static field $\phi = \phi(r)$, we define the auxiliary function $Q$:
\begin{equation}
\label{Def:Q}
Q(r) = \sqrt{1 + \frac{2X}{\Lambda^4}} = \sqrt{1 - \frac{f(r)(\phi')^2}{\Lambda^4}}.
\end{equation}
The Energy Momentum tensor corresponding to the generic DBI matter Lagrangian is given in the RHS of \eqref{SM:DBI:metric:Eom}. We express the components of the Energy Momentum tensor in terms of the auxiliary function $Q(r)$ defined above:  \begin{align}
T^t_t &= V(\phi) (1 - Q), \\
T^r_r &= V(\phi) \left( 1 - \frac{1}{Q} \right).
\end{align}
We express the metric field equations again in terms of the auxiliary function $Q(r)$ as follows:
\begin{eqnarray}
    G^{t}_{t}=\kappa^2V(\phi)\left(1-Q(r)\right) \\
    G^{t}_{t}-G^{r}_{r}=\kappa^{2}V(\phi)\left(\frac{1}{Q}-Q\right)
\end{eqnarray}
Taking the ratio of the second equation to the first eliminates $V$:
\begin{equation}
\frac{G^t_t - G^r_r}{G^t_t} = \frac{ \kappa^2 V \frac{(1-Q^2)}{Q}}{ \kappa^2 V (1-Q)} =  \frac{1+Q}{Q} = \left( \frac{1}{Q} + 1 \right).
\end{equation}
Solving for $Q(r)$, we obtain the following simple expression:
\begin{equation}
\label{finalexp:Q}
Q(r) = -\frac{G^t_t}{G^r_r}.
\end{equation}
From \eqref{Def:Q}, we obtain: 
\begin{eqnarray}
    \frac{f(r)(\phi^{\prime}(r))^2}{\Lambda^4} =1-Q^{2}(r)
\end{eqnarray}
Using \eqref{finalexp:Q}, we obtain the following expression:
\begin{eqnarray}
    \label{exp:phi}
    \phi(r)=\pm\int  dr\sqrt{\frac{\Lambda^4}{f(r)}\left(1-\frac{(G^{t}_{t})^2}{(G^{r}_{r})^{2}}\right)}
\end{eqnarray}
The RHS of the above equation can be completely determined from the geometry. Similarly, we can also express the potential $V(\phi(r))$ in terms of the geometry in the following form:
    \begin{eqnarray}
       V(\phi(r))=\frac{G^{t}_{t}}{\kappa^2\Lambda^4(1-Q(r))} =\frac{G^{t}_{t}\hspace{0.1cm}G^{r}_{r}}{\kappa^2\Lambda^{4}(G^{t}_{t}+G^{r}_{r})}
    \end{eqnarray}

For a reference, we write the full expressions of $Q(r)$ and $V(r)$ below:
\begin{equation}
    \label{exp:Q}
    Q(r)=\frac{-3 GM \left(a^2+r^2\right) \left(2 a^2+3 r^2\right) \tan ^{-1}\left(\frac{a}{r}\right)+12 a^3 GM r + a^5+9 a {GM} r^3}{9 {GM} r^2 \left(a^2+r^2\right) \tan ^{-1}\left(\frac{a}{r}\right)-6 a^3 {GM}r + a^5-9 a {GM} r^3}
\end{equation}

\begin{equation}
     \label{exp:V}
     V(r)=\frac{-3 {GM} \left(a^2+r^2\right) \left(2 a^2+3 r^2\right) \tan ^{-1}\left(\frac{a}{r}\right)+12 a^3 {GM} r+a^5+9 a {GM} r^3}{a^3 \kappa ^2 \Lambda ^4 \left(a^2+r^2\right)^2 \left(1-\frac{-3 {GM} \left(a^2+r^2\right) \left(2 a^2+3 r^2\right) \tan ^{-1}\left(\frac{a}{r}\right)+12 a^3 {GM} r+a^5+9 a {GM} r^3}{9 {GM} r^2 \left(a^2+r^2\right) \tan ^{-1}\left(\frac{a}{r}\right)-6 a^3 {GM} r+a^5-9 a {GM} r^3}\right)}
 \end{equation}
For a consistency check, we consider the EOM satisfied by the DBI scalar field $\phi(r)$.
In the metric ansatz \eqref{SM:metric:ansatz}, the scalar eom given in \eqref{SM:Phi:EOM}, takes the following form:
 \begin{equation}
     \frac{d}{dr}\left(\frac{\rho^2f(r)\phi^{\prime}V(\phi(r))}{Q(r)}\right)=\rho^2\Lambda^4\frac{dV}{d\phi}\left(1-Q(r)\right)
 \end{equation}
After substituting \eqref{exp:Q}, \eqref{exp:V}, \eqref{exp:phi}, the above EOM for $\phi$ is automatically satisfied.
\par{} 
For another consistency check, we look at the asymptotic behaviour of $\phi^{\prime}(r)$. We know that
\begin{equation}
    (\phi^{\prime}(r))^2=\frac{\Lambda^4}{f(r)}(1-Q^{2}(r))
\end{equation}
The behaviour of $Q(r)^2$ in the asymptotic limit is given as
\begin{equation}
    \lim_{r\rightarrow\infty} Q^{2}(r)=1-\frac{16 GM}{5r}+\mathcal{O}\left(\frac{1}{r^2}\right)
\end{equation}

\begin{eqnarray}
    \lim_{r\rightarrow\infty}\frac{(1-Q^{2}(r))}{f(r)}=\frac{16GM}{5r}+\frac{192(GM)^{2}}{25r^2}+\mathcal{O}\left(\frac{1}{r^3}\right)
\end{eqnarray}

Hence, we find
\begin{equation}
\lim_{r\rightarrow\infty}\phi^{\prime}(r)=0
\end{equation} 
Hence, we find that in the limit $r\rightarrow\infty$, the Energy Momentum Tensor vanishes which is consistent with the fact the spacetime is asymptotically flat.

\section{Sign of $\epsilon$}
In this section, we fix the sign of $\epsilon$. We use the relation: 
\begin{equation}
    G^{t}_{t}-G^{r}_{r}=\epsilon\kappa^2\left(T^{t}_{t}-T^{r}_{r}\right)
\end{equation}
We can calculate the expression of the LHS from the geometry:
\begin{equation}
    \label{LHS:expression}
    G^{t}_{t}-G^{r}_{r}=\frac{2\rho^{\prime\prime}(r)f(r)}{\rho(r)}
\end{equation}
From the matter sector, the expression of the RHS is of the following form:
\begin{equation}
    \label{RHS:expression}
    \epsilon\kappa^2(T^{t}_{t}-T^{r}_{r})=\frac{\epsilon\kappa^2V(\phi(r))f(r)(\phi^{\prime}(r))^2}{\Lambda^4Q(r)}
\end{equation}
Hence, we obtain
\begin{equation}
    \frac{2\rho^{\prime\prime}(r)}{\rho(r)}=\frac{\epsilon\kappa^2V(\phi(r))(\phi^{\prime})^2}{\Lambda^4 Q(r)}
\end{equation}
\noindent For $\rho=\sqrt{r^2+a^2}$, the LHS takes the form: $\frac{a^2}{(a^2+r^2)^2}$ which is strictly positive.
\vspace{0.1cm}
\\ 
We plot $V(\phi(r))/Q(r)$ below:

\begin{figure}[H]
  \centering
  \begin{subfigure}{0.42\textwidth}
    \centering
    \includegraphics[width=\linewidth]{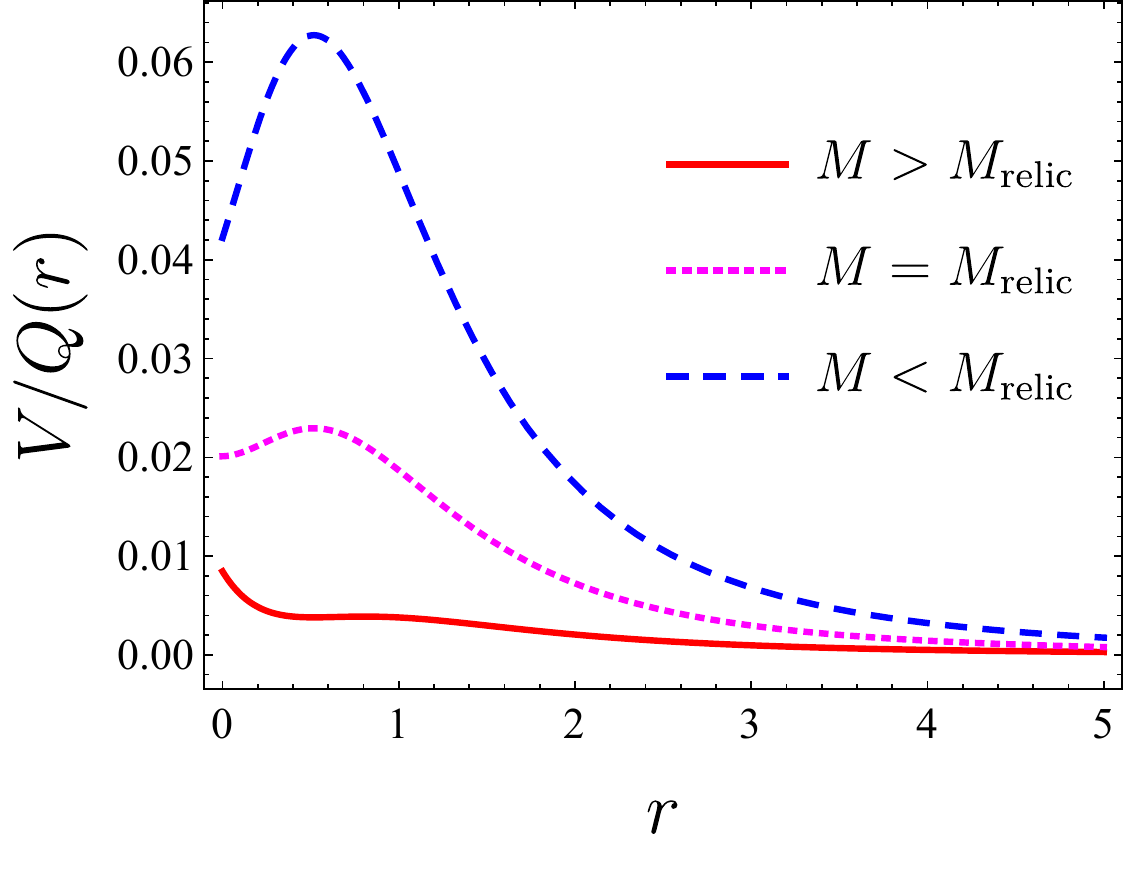}
    \caption{$V/Q(r)$ for the three mass regimes (in units where $a=G=1$) near the core region.}
  \end{subfigure}
  \hfill
  \begin{subfigure}{0.44\textwidth}
    \centering
    \includegraphics[width=\linewidth]{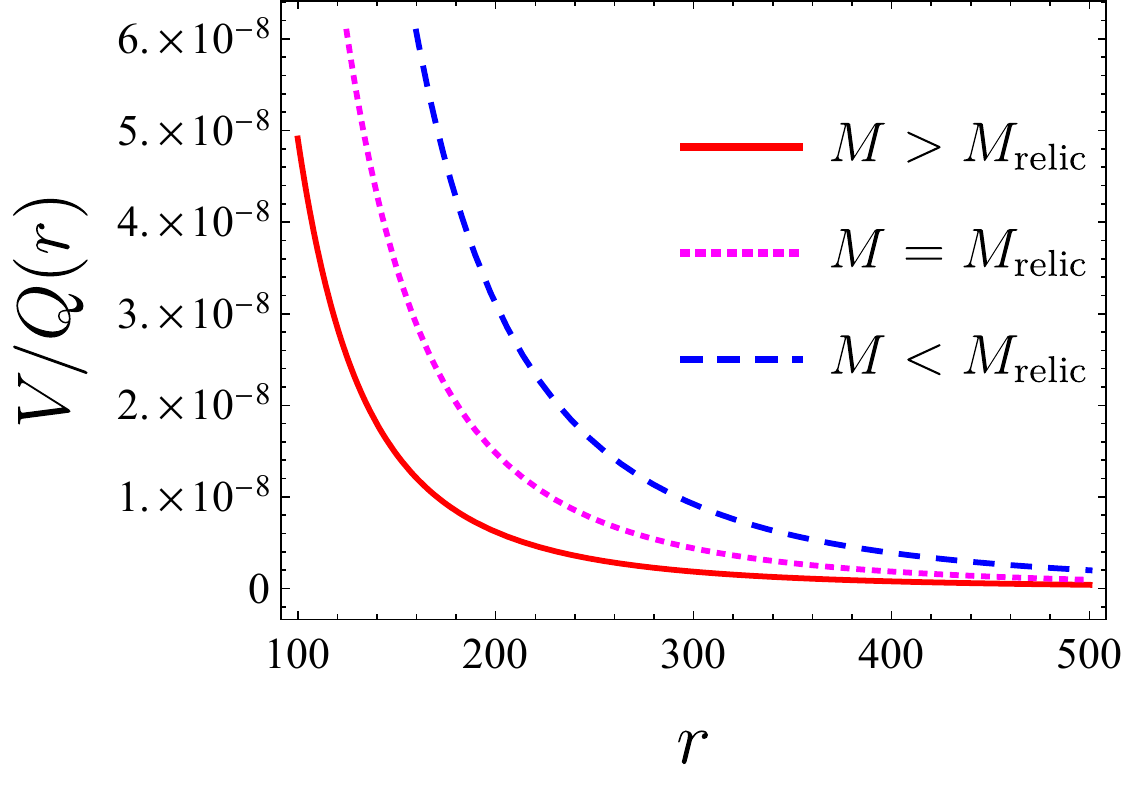}
    \caption{$V/Q(r)$ for the three mass regimes (in units where $a=G=1$) in the large $r$ limit.}
  \end{subfigure}
  \caption{Plot of $V/Q(r)$ for the three mass regimes (in units where $a=G=1$), showing the non-extremal BH (red, $M=0.50$), the stable extremal relic (magenta, $M=2/3\pi$), and the horizon-less regular object (blue, $M=0.10$).}
\end{figure}

\noindent \par{}
From the plot of $V(\phi(r))/Q(r)$, we infer that the ratio is also strictly positive and hence, the equality holds only when $\epsilon=+1$.

\section{{Reality Condition of the scalar field $\phi(r)$}}
In this section, we look at the nature of the DBI scalar field $\phi(r)$. From the integral expression of $\phi(r)$ given in \eqref{exp:phi}, we infer that as longs as the quantity inside the square root is positive, $\phi(r)$ remains real. We define the quantity inside the square root as:
\begin{eqnarray}
    g(r)=\frac{\left(1-Q^{2}(r)\right)}{f(r)}\hspace{0.1cm};\hspace{0.1cm}Q(r)=-\frac{G^{t}_{t}}{G^{r}_{r}}~.
\end{eqnarray}

\begin{figure}[H]
  \centering
  \begin{subfigure}{0.45\textwidth}
    \centering
    \includegraphics[width=\linewidth]{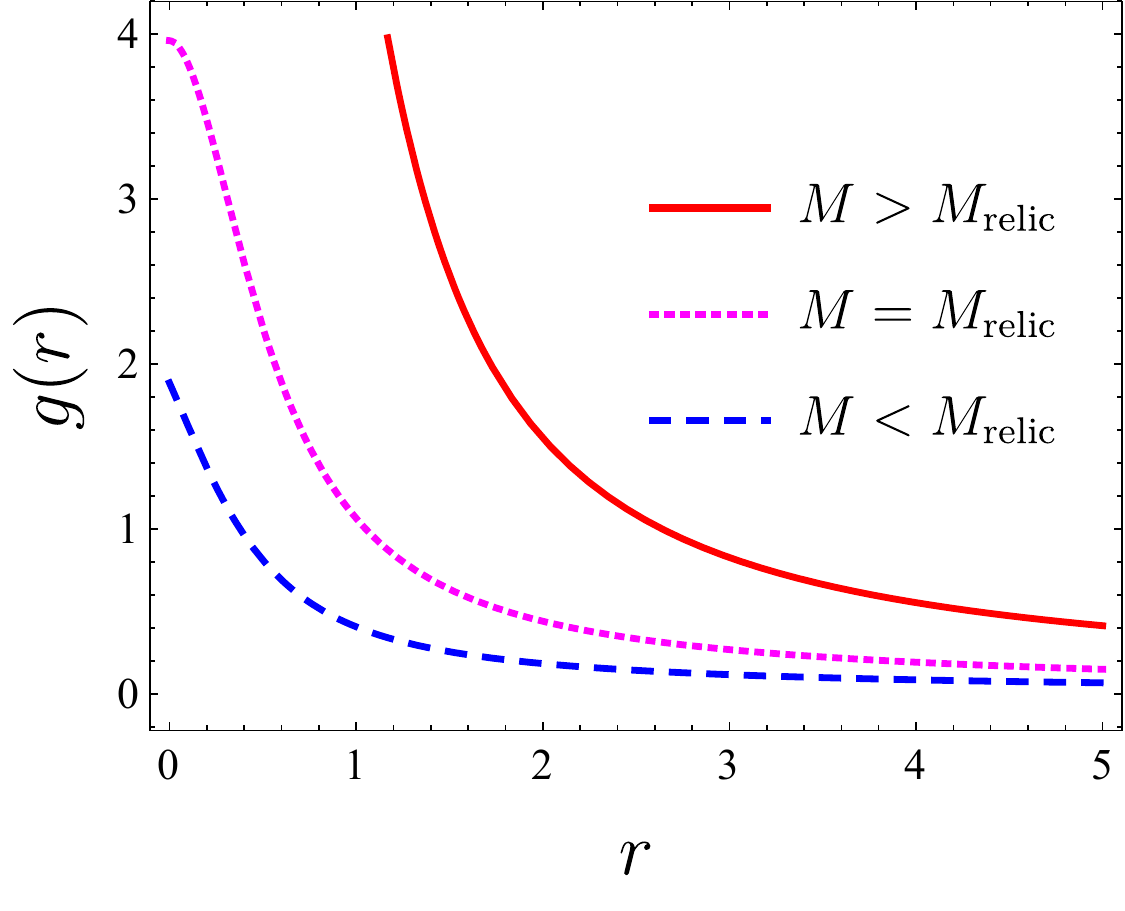}
    \caption{ Plot of $g(r)$ for the three mass regimes (in units where $a=G=1$) near the core region.}
  \end{subfigure}
  \hfill
  \begin{subfigure}{0.49\textwidth}
    \centering
    \includegraphics[width=\linewidth]{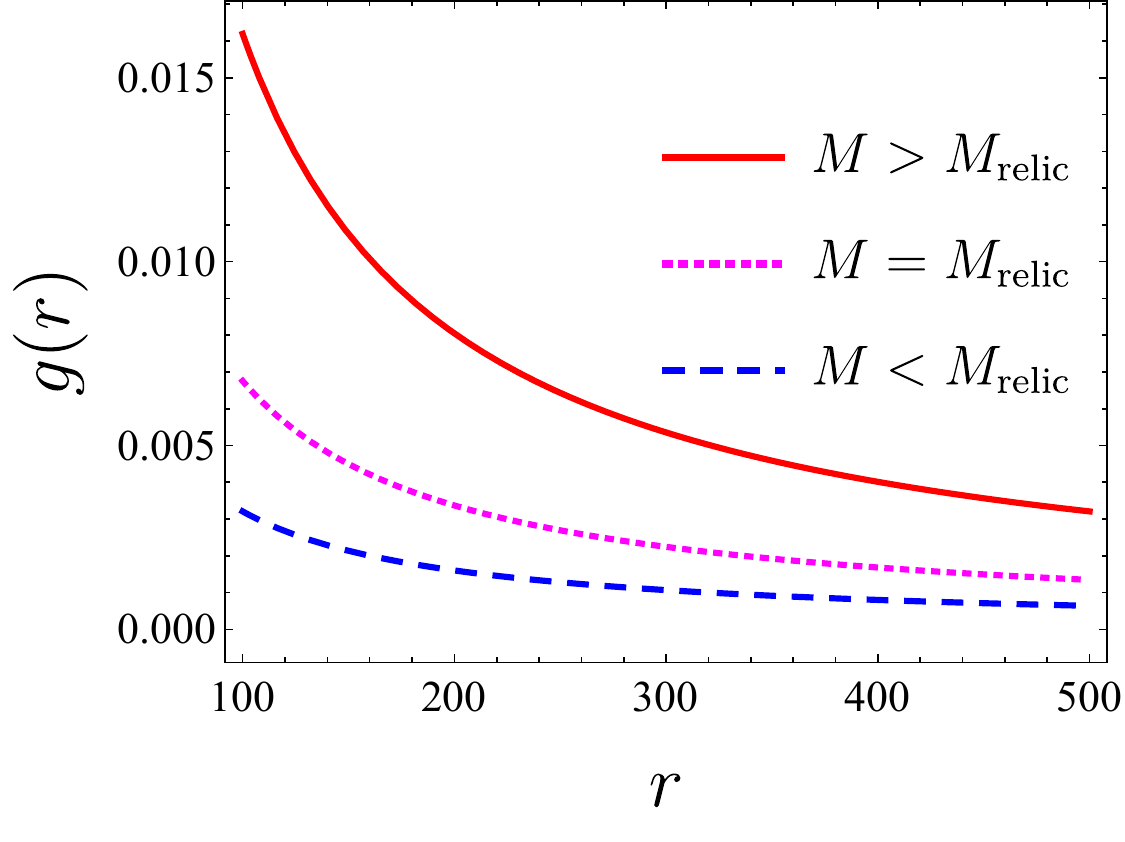}
    \caption{Plot of $g(r)$ for the three mass regimes (in units where $a=G=1$) in the large $r$ limit.}
  \end{subfigure}
\end{figure}
\noindent We plot the function $g(r)$ below for the three different cases (Non-extremal BH, Extremal BH, Regular Object).
From the above plots, we infer that $g(r)$ remains positive both near the core region as well as in the large $r$ limit for the chosen parameters and hence the DBI scalar field $\phi(r)$ remains real.
\\
\section{Curvature invariants and Geodesic completeness}
\label{Invariants_Completeness}
Having solved for the metric function $f(r)$ and scalar field $\phi(r)$, we explicitly compute various curvature invariants and obtain the asymptotic limits.  
\subsection{Ricci scalar ($R$)}
The expression of the Ricci scalar is as follows
\begin{align}
R=\frac{2}{a^3 (a^2+r^2)^2} \Big[ & 9 G M \left(a^2+r^2\right) \left(a^2+2 r^2\right) \tan ^{-1}\left(\frac{a}{r}\right) \nonumber \\
& -21 a^3 G M r-a^5-18 a G M r^3 \Big]~.
\end{align}
To check the regularity of the Ricci scalar at $r=0$, we evaluate the above expression at $r=0$. 
\begin{equation}
\lim_{r\rightarrow 0}R=\frac{9 \pi G M-2 a}{a^3}~.
\end{equation}
From the above expression, we infer that the Ricci scalar is finite and non-divergent in the $r\rightarrow 0$ limit. As expected the value of Ricci scalar depends on the ratio $9 \pi GM/(2 a)$. For completeness, we also look at the asymptotic behaviour of the Ricci scalar.
\begin{equation}
\lim_{r\rightarrow \infty} R= -\frac{2 a^2}{r^4}+\frac{ \left(36 a^2 GM\right)}{5 r^5}+\mathcal{O}\left(\frac{1}{r^6}\right)~.
\end{equation}
From the above expression, we also infer that the Ricci scalar asymptotically approaches $0$ implying that the spacetime is asymptotically flat.

\subsection{Ricci tensor squared ($R^{\mu\nu}R_{\mu\nu}$)}
We write the expression for the square of the Ricci tensor, $R^{\mu\nu}R_{\mu\nu}$.
\begin{align}
& R^{\mu\nu}R_{\mu\nu}=\frac{4}{a^6 (a^2 + r^2)^4} \Bigg[ \,
a^2 \Big(
189 a^2 G^2 M^2 r^4 \nonumber \\
& \quad + 9 a^4 G M r^2 (13 G M + r)
+ 15 a^6 G M r
+ a^8
+ 81 G^2 M^2 r^6
\Big) \nonumber \\
& \quad + 9 G M (a^2 + r^2)^2 
\tan^{-1}\!\left(\frac{a}{r}\right)
\Big(
3 G M (3 a^2 r^2 + a^4 + 3 r^4) \nonumber \\
& \quad \times \tan^{-1}\!\left(\frac{a}{r}\right)
- 9 GMa(a^2+r^2) \Big) \nonumber \\
& \quad - 9 GMa(a^2+r^2)\tan^{-1}\left(\frac{a}{r}\right) ( 12 a^2 G M r + a^4 + 18 G M r^3 )
\Bigg]~.
\end{align}
To check the regularity of $R^{\mu\nu}R_{\mu\nu}$ at $r=0$, we evaluate the expression at $r=0$.
\begin{equation}
\lim_{r\rightarrow 0}R^{\mu\nu}R_{\mu\nu}= \frac{4 a^2-18 \pi a G M+27 \pi ^2 G^2 M^2}{a^6}~.
\end{equation}
From the above expression, we again infer that $R^{\mu\nu}R_{\mu\nu}$ is regular as $r\rightarrow 0$.  Here again, as expected the value of Ricci scalar depends on the ratio $9 \pi GM/(2 a)$.

We also look at the asymptotic behaviour of $R^{\mu\nu}R_{\mu\nu}$.
\begin{equation}
\lim_{r\rightarrow \infty}R^{\mu\nu}R_{\mu\nu}= \frac{4 a^4}{r^8}-\frac{96 \left(a^4 G M\right)}{5 r^9}+\mathcal{O}\left(\frac{1}{r^{10}}\right)~.
\end{equation}
$R^{\mu\nu}R_{\mu\nu}$ asymptotically goes to $0$ as the spacetime is asymptotically flat.

\subsection{Kretschmann scalar ($K$)}
We also write the expression of the Kretschmann scalar $K=R_{\mu\nu\alpha\beta}R^{\mu\nu\alpha\beta}$.
\begin{align}
K = & \frac{12}{a^6 (a^2 + r^2)^4} \Bigg[ \,
a^2 \Big(
42 a^2 G^2 M^2 r^4 \nonumber \\
& \quad + a^4 G M r^2 (33 G M + 2 r)+ 8 a^6 G M r
+ a^8 \nonumber \\
& \quad + 18 G^2 M^2 r^6\Big) 
+ 9 G^2 M^2 (a^2 + r^2)^2 (2 a^2 r^2 + a^4 + 2 r^4) \nonumber \\
& \quad \times \left[\tan^{-1}\!\left(\frac{a}{r}\right)\right]^2 
- 2 a G M (a^2 + r^2)
\tan^{-1}\!\left(\frac{a}{r}\right) \nonumber \\
& \quad \times \Big(
30 a^2 G M r^3+a^{4}r(15GM+r)
+2a^6+18GMr^5\Big)
\Bigg]~.
\end{align}
To check the regularity of $K$ at $r=0$, we evaluate the expression at $r=0$.
\begin{equation}
\label{K_scalar_at_r=0}
\lim_{r\rightarrow 0 }K = \frac{3 \left(4 a^2-8 \pi a G M+9 \pi ^2 G^2 M^2\right)}{a^6}~.
\end{equation}
The Kretschmann scalar is also regular at $r=0$. We also look at the asymptotic behaviour of the Kretschmann scalar.
\begin{equation}
\label{K_scalar_at_infinity}
\lim_{r\rightarrow\infty} K= \frac{48 G^2 M^2}{r^6}+\frac{32 a^2 G M}{r^7}+\mathcal{O}\left(\frac{1}{r^8}\right)~.
\end{equation}
Like the previous two invariant scalar quantities, the Kretschmann scalar also approaches $0$ asymptotically as the spacetime approaches flat geometry.

\begin{figure}[t]
 \centering
    \includegraphics[width=0.55\linewidth]{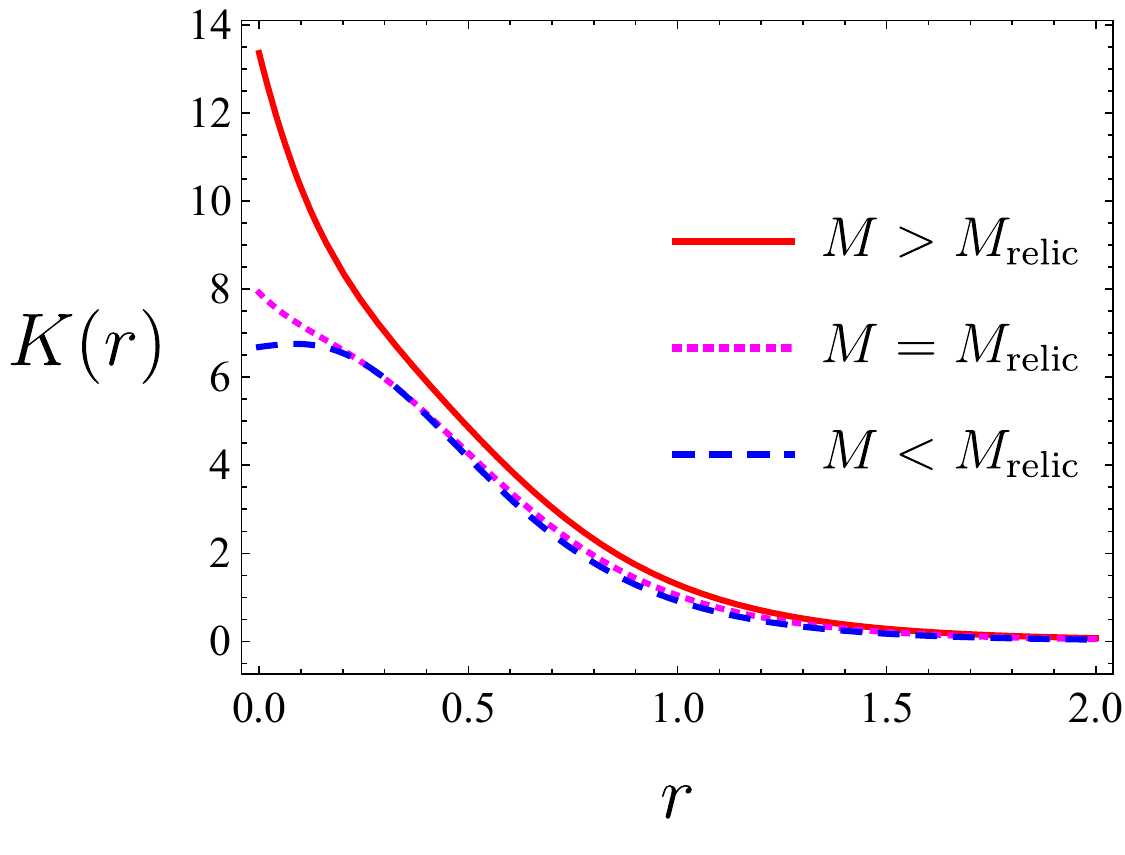}
\caption{The Kretschmann scalar $K(r)$ for the three mass regimes (in units where $a=G=1$), showing the non-extremal BH (red, $M=0.30$), the stable extremal relic (magenta, $M=2/3\pi$), and the horizon-less regular object (blue, $M=0.15$).}
\label{SM:fig:K_plot}
\end{figure}

Figure \ref{SM:fig:K_plot} shows the behavior of the Kretschmann scalar $K(r)$ for the three mass regimes identified in the main text. From this solution, we infer two key properties:
\begin{enumerate}
    \item From \eqref{K_scalar_at_r=0}, we see that the value of the Kretschmann scalar increases with $M$ in the $r\rightarrow 0$ limit. Hence, the value of the Kretschmann scalar is large for the extremal Black Hole $(M>M_{relic})$ compared to the case of the extremal Black Hole $(M=M_{relic})$ and the horizon-less regular object $M<M_{relic}$ in the $r\rightarrow0$ limit. 
    \item For all mass regimes, the Kretschmann scalar asymptotically approaches $0$ as expected from \eqref{K_scalar_at_infinity}.
\end{enumerate}

\subsection{Geodesic Completeness}
\label{sec:geodesics}
A spacetime is non-singular if and only if all causal (timelike and null) geodesics are complete. We analyze the radial null geodesics, which are the most direct probes of the spacetime's causal structure.
For a radial null geodesic ($ds^2=0, d\theta^2=d\phi^2=0$), the metric \eqref{SM:metric:ansatz} gives:
\begin{equation}
    f(r) dt^2 = \frac{dr^2}{f(r)} \implies \frac{dt}{dr} = \pm \frac{1}{f(r)}
\end{equation}
The key to geodesic completeness is to analyze the affine parameter $\lambda$. For a null geodesic $k^\mu = dx^\mu / d\lambda$, the radial equation is:
\begin{equation}
    g_{\mu\nu}k^\mu k^\nu = -f(r)(k^t)^2 + \frac{1}{f(r)}(k^r)^2 = 0 \implies k^r = \pm f(r) k^t
\end{equation}
From the Killing vector $\partial_t$, we have a conserved quantity, the energy $E$:
\begin{equation}
    E = -g_{\mu\nu} (\partial_t)^\mu k^\nu = -g_{tt} k^t = f(r) k^t \implies k^t = \frac{E}{f(r)}
\end{equation}
Substituting this back into the $k^r$ equation:
\begin{equation}
    k^r = \frac{dr}{d\lambda} = \pm f(r) \left(\frac{E}{f(r)}\right) = \pm E
\end{equation}
 To find the total affine parameter $\Delta \lambda$ required to reach the central region $r=0$ from some initial radius $r_0$, we integrate:
\begin{equation}
    \Delta \lambda = \int d\lambda = \int_{r_0}^{0} \frac{d\lambda}{dr} dr = \int_{r_0}^{0} \frac{dr}{(\pm E)} = \mp \frac{r_0}{E}
\end{equation}
The affine parameter $\Delta \lambda$ is \emph{finite}. This means an observer on a radial null geodesic reaches the center $r=0$ in a finite affine time. In our regular spacetime, $r=0$ is just another point, a regular 2-sphere. The geodesic is not terminated; it simply passes \emph{through} $r=0$ and can be extended to negative $r$ values, effectively emerging into a new asymptotically flat region (like a wormhole throat).

Since the affine parameter is finite and the geodesic can be continued indefinitely (from $r=-\infty$ to $r=+\infty$), all radial null geodesics are complete. A similar analysis shows that timelike geodesics are also complete. Therefore, the spacetime is geodesically complete and free of singularities.

\section{Energy Condition: Sign of $\mathbf{T^{t}_{t}}$}

A critical check for any classical solution is its adherence to fundamental energy conditions. The Weak Energy Condition (WEC) states that the energy density $\rho$ must be non-negative for any timelike observer. For our static, spherically symmetric spacetime, the energy density measured by an observer at rest is $\rho = -T^t_t$. We therefore investigate the sign of $T^t_t$ to determine where, if at all, the WEC is satisfied.

From the field equations \eqref{SM:DBI:metric:Eom}, the $T^t_t$ component of the stress-energy tensor is:
\begin{equation}
\label{SM:Ttt:exp}
T^{t}_{t}=\epsilon\hspace{0.1cm} 
V(\phi)\bigg(1-\sqrt{1+\frac{2X}{\Lambda^4}}\bigg)~.
\end{equation}
Using $X = -f(r)(\phi')^2/2$, this becomes:
\begin{equation}
\label{SM:Ttt:exp2}
T^{t}_{t}=\epsilon\hspace{0.1cm}
V(\phi)\bigg(1-\sqrt{1-\frac{f(r)(\phi^{\prime}(r))^2}{\Lambda^4}}\bigg)~.
\end{equation}
We plot the function $T^{t}_{t}(r)$ for the three different mass regimes below:
\begin{figure}[H]
 \centering
    \includegraphics[width=0.60\linewidth]{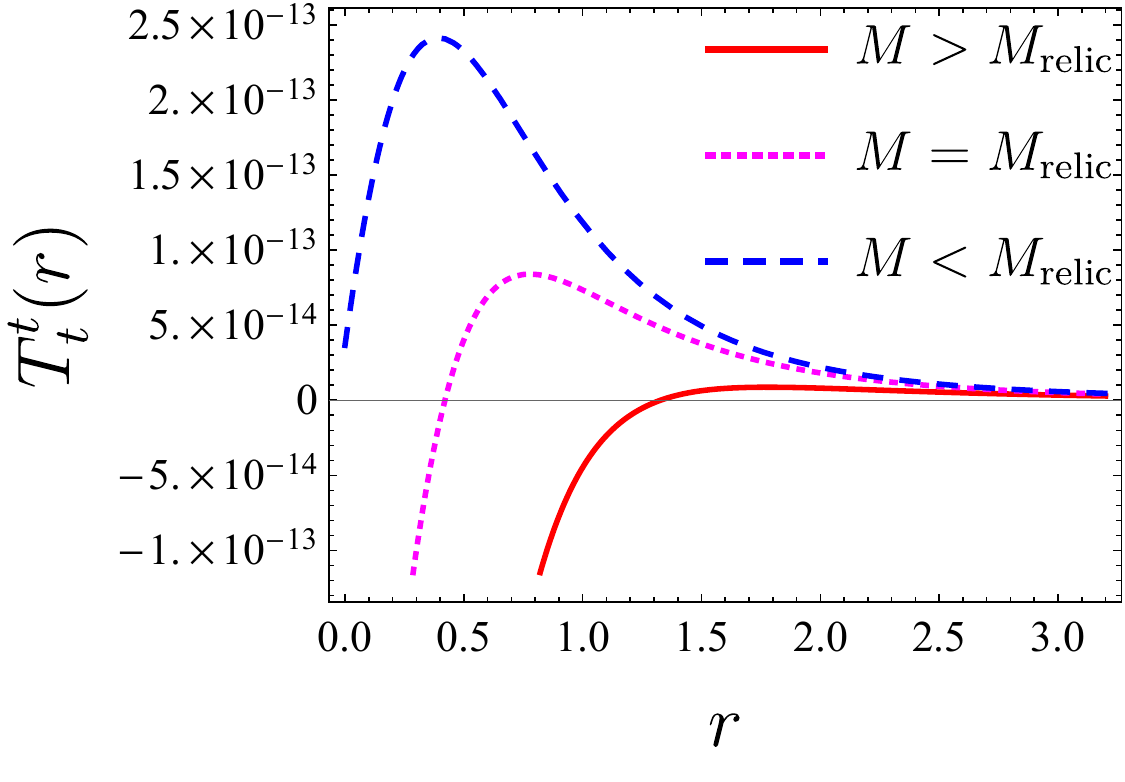}
\caption{$T^{t}_{t}(r)$ for the three mass regimes (in units where $a=G=1$), showing the non-extremal BH (red, $M=0.50$), the stable extremal relic (magenta, $M=2/3\pi$), and the horizon-less regular object (blue, $M=0.10$).}
\label{SM:fig:K_plot}
\end{figure}

We infer about the energy condition from the sign of $T^{t}_{t}$ or $\rho$ (for $\epsilon=1$).
{We make the following conclusions for the three mass regimes:
\begin{itemize}
    \item \textbf{Horizon-less Object ($M < M_{\text{relic}}$):} From the plot of $T^{t}_{t}$, we infer that $T^{t}_{t}>0$ and hence the energy density $\rho<0$ for all radii. The WEC is violated at all radii when $M<M_\text{relic}$.
    \item \textbf{Extremal Relic ($M = M_{\text{relic}}$):} From the plot of $T^{t}_{t}$, we infer that near the core region, $T^{t}_{t}<0$ and hence $\rho>0$. Far outside the horizon, $T^{t}_{t}>0$ and hence $\rho<0$. Hence, we conclude that the WEC is violated far outside the horizon.
    \item \textbf{Non-extremal BH ($M > M_{\text{relic}}$):} Ftom the plot of $T^{t}_{t}$, we infer that near the core region, $T^{t}_{t}<0$ and hnece $\rho>0$. Far outside the horizon, $T^{t}_{t}>0$ and hence $\rho<0$. Hence, we conclude that the WEC is violated far outside the horizon.
\end{itemize}}
This violation of the WEC is a common feature of non-singular black hole models and is directly responsible for evading the Penrose-Hawking singularity theorems.
\section{Relation between the parameters $a,M,\Lambda$}
\label{relation_between_parameters}
We obtain a relation between the parameters $a,M,\Lambda$. $\Lambda$ is the energy scale associated with the DBI field. $M$ appears as an integration constant and $a$ is introduced in the ansatz of the areal radius requiring a regular solution. We equate the core energy density with the energy scale ($\Lambda$) of the theory. The expression of $T^{t}_{t}$ is given below:
\begin{equation}
    T^{t}_{t}(r)=\frac{-3 {GM} \left(a^2+r^2\right) \left(2 a^2+3 r^2\right) \left(\frac{\pi }{2}-\tan ^{-1}\left(\frac{r}{a}\right)\right)+12 a^3 {GM} r+a^5+9 a {GM} r^3}{a^3 \kappa ^2 \left(a^2+r^2\right)^2}~.
\end{equation}
Now, we equate the energy density ($\rho$) in the core region ($r=0$) with the energy scale $\Lambda$.
\begin{eqnarray}
    \rho_{\text{core}}=-T^{t}_{t}(r=0)=\frac{3\pi GM-a}{a^3\kappa^2}~.
\end{eqnarray}
For BHs satisfying ($GM>>a$), we obtain the following simple relation:
\begin{eqnarray}
    \label{parameters:relation}
    \frac{3\pi GM}{a^3\kappa^2}\sim \Lambda^4\hspace{0.1cm}\Longrightarrow a\sim \left(\frac{3M}{8\Lambda^4}\right)^{\frac{1}{3}}~.
\end{eqnarray} 
\noindent In the above expression, we have used the relation $\kappa^2=8\pi G$.

\section{Black Hole Thermodynamics}
In this section, we analyze the thermodynamic properties of the non-extremal black hole solution ($M > M_{\text{relic}}$). We focus on the large-mass (small $a$) limit, where $a \ll 2GM$, where classical approximations are most secure. We derive the corrections to the standard Schwarzschild results by expanding our solution in powers of $a$.

First, we expand the metric function $f(r)$ from Eq.~\eqref{SM:f(r):exp}. Using the Taylor series for $\tan^{-1}(x) = x - x^3/3 + x^5/5 - \dots$, we have:
\begin{equation}
\tan^{-1}\left(\frac{a}{r}\right)=\frac{a}{r}-\frac{a^3}{3r^3}+\frac{a^5}{5r^7}-\mathcal{O}\left( a^7\right)~.
\end{equation}
We substitute this into \eqref{SM:f(r):exp} and expand:
\begin{align}
f(r) &=1+\frac{3GM}{a}\bigg[\frac{r}{a}-\frac{a^2+r^2}{a^2}\left(\frac{a}{r}-\frac{a^3}{3r^3}+\dots\right)\bigg] \nonumber \\
& =1-\frac{2GM}{r}+\frac{2a^2GM}{5r^3}+\mathcal{O}(a^4)~.
\end{align}
This confirms that our solution is the Schwarzschild metric plus a small $a^2$ correction term.

\subsection{Horizon Radius}
To calculate the Hawking temperature, we must first find the precise location of the event horizon $r_h$ by solving $f(r_h) = 0$.
\begin{equation}
1-\frac{2GM}{r_{h}}+\frac{2GMa^2}{5r^{3}_{h}}\approx 0~.
\end{equation}
We assume a perturbative ansatz $r_{h}=2GM+\delta $, where $\delta$ is a small correction of order $\mathcal{O}(a^2)$. Substituting this into the horizon condition:
\begin{align}
& 1-\frac{2GM}{2GM+\delta}+\frac{2GMa^2}{5(2GM+\delta)^3}\approx 0 \nonumber \nonumber \\
& \implies \frac{\delta}{2GM} \approx -\frac{a^2}{20G^2M^2}~~{\rm or}~~\delta = -\frac{a^2}{10GM}~.
\end{align}
The corrected horizon radius is therefore slightly smaller than the Schwarzschild radius:
\begin{equation}
r_{h} \approx 2GM-\frac{a^2}{10GM}~.
\end{equation}

\subsection{Hawking Temperature}
The Hawking temperature is defined by the surface gravity $\kappa_s$ at the horizon:
\begin{equation}
T_{H}=\frac{\kappa_{s}}{2\pi}=\frac{f^{\prime}(r_{h})}{4\pi}~.
\end{equation}
First, we compute the derivative of the expanded metric function:
\begin{align}
f^{\prime}(r) & \approx \frac{d}{dr}\left(1 - 2GMr^{-1} + \frac{2GMa^2}{5}r^{-3}\right) \\
& = \frac{2GM}{r^2}-\frac{6GMa^2}{5r^4}~.
\end{align}
Now, we evaluate this at the corrected horizon $r_h \approx 2GM$:
\begin{align}
f^{\prime}(r_{h})& \approx \frac{2GM}{(2GM - \frac{a^2}{10GM})^2}-\frac{6GMa^2}{5(2GM)^4} \nonumber \\
& = \frac{1}{2GM} + \frac{a^2}{20G^3M^3} - \frac{3a^2}{40G^3M^3} \nonumber \\
& = \frac{1}{2GM}\left(1-\frac{a^2}{20G^2M^2}\right)~.
\end{align}
The Hawking temperature is therefore:
\begin{align}
T_{H} &= \frac{1}{4\pi}f^{\prime}(r_{h}) =\frac{1}{8\pi GM}\left(1-\frac{a^2}{20(GM)^2}\right) \\
\label{Temp:FinalExp}
T_{H} & =T_{S}\left(1-\frac{a^2}{20(GM)^2}\right)~.
\end{align}
where $T_{S} = 1/(8\pi GM)$ is the standard Schwarzschild temperature. This shows our black hole is slightly cooler than a Schwarzschild black hole of the same mass.

\subsection{Evaporation rate}
The evaporation rate (Power, $P$) is set by the Stefan-Boltzmann law, $P \propto A_{h} T_{H}^4$. The area of the horizon $A_h$ is given by the area of the minimal sphere at $r_h$, which is $A_h = 4\pi \rho(r_h)^2 = 4\pi (r_h^2+a^2)$.
\begin{align}
A_{h} &= 4\pi \left(\left(2GM-\frac{a^2}{10GM}\right)^2+a^2\right) 
\nonumber \\
 & = A_{S}\bigg(1+\frac{3a^2}{20(GM)^2}\bigg)~.
\end{align}
where $A_{S} = 16\pi G^2 M^2$ is the Schwarzschild area.
The evaporation rate is then:
\begin{align}
P & \propto A_{h}(T_{H})^4  \nonumber \\
& \propto \bigg[A_{S}\left(1+\frac{3a^2}{20(GM)^2}\right)\bigg]\bigg[T^4_{S}\left(1-\frac{a^2}{20(GM)^2}\right)^4\bigg] \nonumber \\
P & \approx P_{S}\bigg(1-\frac{a^2}{20(GM)^2}\bigg)~.
\end{align}
where $P_S \propto A_S T_S^4$ is the Schwarzschild rate. The evaporation rate is slightly slower than that of a Schwarzschild black hole, implying a slightly longer lifetime. As discussed in the main text, this perturbative-regime calculation gives way to the non-perturbative result that $T_H \to 0$ and evaporation ceases entirely as $M \to M_{\text{relic}}$.
\vspace{0.1cm}
\par{}
We can obtain $M_{\text{relic}}$ from \eqref{Temp:FinalExp}. Setting the expression in \eqref{Temp:FinalExp} equal to $0$, we obtain:


\begin{eqnarray}
    \frac{a^2}{20}=(GM_{\text{relic}})^2
\end{eqnarray}

Using \eqref{parameters:relation}, we obtain:

\begin{eqnarray}
    \left(M_{\text{relic}}\right)^{\frac{4}{3}}= \left(\frac{3}{8}\right)^{\frac{2}{3}}\times \left(\frac{64\pi^2}{20}\right)\times\left(\frac{1}{\kappa^4(\Lambda)^{\frac{8}{3}}}\right)
\end{eqnarray}

Hence, using the fact $\Lambda \sim 10^{17}$ GeV and $M_{\text{Planck}}\sim10^{19}$ GeV, we obtain
\begin{eqnarray}
    M_{\text{relic}}\sim 8\times 10^{4}\times M_{\text{Planck}}
\end{eqnarray}
Hence, finally, we obtain
\begin{eqnarray}
    M_{\text{Planck}}\sim 0.8 gm
\end{eqnarray}

\section{Phantom field from Non-linear sigma model}
\label{phantom_DBI}
In modern cosmology and high-energy physics, complex scalar field Lagrangians, such as the Dirac-Born-Infeld (DBI) or k-essence models, are frequently employed \cite{Armendariz-Picon:1999hyi}. Often, these non-canonical terms are not fundamental but emerge as the low-energy effective action of a more complex, multi-field theory.

A common example of this is a Non-Linear Sigma Model (NLSM), which describes the dynamics of Goldstone bosons arising from spontaneous symmetry breaking~\cite{Callan:1969sn, Weinberg:1978kz}. In these models, one can ``integrate out" heavy or auxiliary fields to find the effective action for the remaining light, dynamical fields. This process can generate highly non-trivial kinetic structures\footnote{We thank Susobhan Mandal for making this point to us.}.

Here, we demonstrate how a simple two-field NLSM, with one dynamical field ($\phi$) and one auxiliary field ($\chi$), can be reduced to an effective single-field theory for $\phi$ that is non-canonical and exhibits phantom behavior.

We begin with the starting Lagrangian for the two scalar fields $\phi$ and $\chi$. The field $\phi$ has a canonical kinetic term $X = -\frac{1}{2}g^{\mu\nu}\nabla_\mu\phi\nabla_\nu\phi$, while $\chi$ is an auxiliary field with no kinetic term:
\begin{equation}
    \mathcal{L}(\phi, \chi) = \left(1 + \frac{X}{\Lambda^4}\right) \chi^a + \chi^b
\end{equation}
where $a$ and $b$ are arbitrary powers, and $\Lambda$ is the energy scale of the theory.

Since $\chi$ is non-dynamical, its equation of motion is a simple algebraic constraint. We find it by varying the Lagrangian with respect to $\chi$:
\begin{equation}
    \frac{\partial \mathcal{L}}{\partial \chi} = a \chi^{a-1} \left(1 + \frac{X}{\Lambda^4}\right) + b \chi^{b-1} = 0
\end{equation}
We can solve this constraint equation for $\chi$ in terms of $X$:
\begin{align}
    \chi^{a-b} &= -\frac{b}{a} \left(1 + \frac{X}{\Lambda^4}\right)^{-1}
\end{align}
This gives the ``on-shell" solution for the $\chi$ field:
\begin{equation}
    \label{eq:chi_solution}
    \chi(X) = \left[ -\frac{b}{a} \right]^{\frac{1}{a-b}} \left(1 + \frac{X}{\Lambda^4}\right)^{-\frac{1}{a-b}}
\end{equation}
Now, we ``integrate out" $\chi$ by substituting this solution back into the original Lagrangian to obtain the effective Lagrangian for $\phi$, $\mathcal{L}_{\text{eff}}(\phi)$:
\begin{align}
    \mathcal{L}_{\text{eff}} &= \left(1 + \frac{X}{\Lambda^4}\right) \chi^a + \chi^b \nonumber \\
    &= \left(1 - \frac{a}{b}\right) \left(1 + \frac{X}{\Lambda^4}\right) \chi^a
\end{align}
Substituting the expression for $\chi^a = (\chi(X))^a$ from Eq.~\eqref{eq:chi_solution}:
\begin{align}
    \mathcal{L}_{\text{eff}} &= \left(\frac{b-a}{b}\right) \left(1 + \frac{X}{\Lambda^4}\right) \left( \left[ -\frac{b}{a} \right]^{\frac{1}{a-b}} \left(1 + \frac{X}{\Lambda^4}\right)^{-\frac{1}{a-b}} \right)^a \nonumber \\
    &= \left(\frac{b-a}{b}\right) \left[ -\frac{b}{a} \right]^{\frac{a}{a-b}} \left(1 + \frac{X}{\Lambda^4}\right)^{1 - \frac{a}{a-b}} \nonumber \\
    &= C(a,b) \cdot \left(1 + \frac{X}{\Lambda^4}\right)^{-\frac{b}{a-b}}
\end{align}
where $C(a,b)$ is a constant that depends only on $a$ and $b$.

\section*{The Phantom Field Case}

Choosing $b = -a$ we get: 
\begin{equation}
-\frac{b}{a-b} = \frac{1}{2};
\quad C(a, -a) = 2
\end{equation}
Therefore, the effective Lagrangian for $\phi$ is:
\begin{equation}
    \mathcal{L}_{\text{eff}} = 2 \sqrt{1 + \frac{X}{\Lambda^4}}
\end{equation}
Note that this non-canonical action was generated purely from the internal dynamics of the two-field model. Thus, our action Eq.~\eqref{DBI:Lagrangian} can be written as:
\begin{equation}
\label{DBI:Lagrangian2}
S=\int d^4x\sqrt{-g}\bigg[\frac{R}{2\kappa^2}+\epsilon\Lambda^4\left(1-\frac{1}{2} \mathcal{L}_{\text{eff}} \right)\bigg]~,
\end{equation}

The physical nature of this effective field is revealed in the low-energy limit, i.e., for small field gradients $X \ll \Lambda^4$. We can Taylor expand the Lagrangian using $\sqrt{1+z} \approx 1 + z/2 + \mathcal{O}(z^2)$:
\begin{equation}
    \mathcal{L}_{\text{eff}} \approx 2 \left( 1 + \frac{1}{2} \frac{X}{\Lambda^4} \right) = 2 + \frac{X}{\Lambda^4}
\end{equation}
The constant `2' acts as a cosmological constant and can be ignored. The kinetic part of the Lagrangian is:
\begin{equation}
    \mathcal{L}_{\text{kin}} \approx +\frac{1}{\Lambda^4} X = -\frac{1}{2\Lambda^4} g^{\mu\nu}\nabla_\mu\phi\nabla_\nu\phi
\end{equation}
A standard canonical scalar field has a Lagrangian $\mathcal{L}_{\text{canonical}} = X$. A field with a ``wrong-sign" kinetic term, $\mathcal{L}_{\text{kin}} \propto -X$, is known as a \emph{phantom field} \cite{Arkani-Hamed:2003pdi}.

Thus, this simple NLSM model, upon integrating out the auxiliary field $\chi$, dynamically generates an effective theory for $\phi$ that is both non-canonical and phantom-like.

\section{Derivation of Scalar Quasi-Normal Modes and Stability Analysis}
\label{QNMs}

In this section, we analyze the linear stability of the black hole solution by computing the Quasi-Normal Modes (QNMs) for a massless scalar perturbation. We consider the background spacetime defined by the line element:
\begin{equation}
    ds^2 = -f(r)dt^2 + \frac{dr^2}{f(r)} + \rho^{2}(r)(d\theta^2 + \sin^2\theta d\phi^2) \,,
\end{equation}
where the metric functions are given by:
\begin{equation}
    f(r) = 1 + \frac{3GM}{a}\left[\frac{r}{a}-\frac{\rho^2(r)}{a^2}\tan^{-1}\left(\frac{a}{r}\right)\right], \quad \rho(r)=\sqrt{r^2+a^2} \,.
\end{equation}
To facilitate the analysis, we introduce the tortoise coordinate $r^*$, defined by the differential relation $dr^{*} = f(r)^{-1} dr$. The line element can then be rewritten in a conformally static form:
\begin{equation}
    ds^2 = f(r)(-dt^2 + dr^{*2}) + \rho^{2}(r)(d\theta^2 + \sin^2\theta d\phi^2) \,.
\end{equation}
The dynamics of a massless scalar field $\Psi$ are governed by the Klein-Gordon equation:
\begin{equation}
    \Box\Psi = \frac{1}{\sqrt{-g}}\partial_{\mu}\left(g^{\mu\nu}\sqrt{-g}\partial_{\nu}\Psi\right) = 0 \,.
\end{equation}
Exploiting the staticity and spherical symmetry of the background, we decompose the scalar field as:
\begin{equation}
    \Psi(t,r,\theta,\phi) = e^{-i\omega t} Y_{lm}(\theta,\phi) \frac{R(r^{*})}{\rho(r)} \,.
\end{equation}
Substituting this ansatz into the wave equation leads to a Schrödinger-like ordinary differential equation for the radial function $R(r^*)$:
\begin{equation}
    \frac{d^2 R(r^{*})}{d{r^{*2}}} + \left(\omega^2 - V_{\text{eff}}(r^*)\right)R(r^{*}) = 0 \,.
\end{equation}
The effective Regge-Wheeler potential $V_{\text{eff}}$ is derived as:
\begin{equation}
    V_{\text{eff}}(r) = f(r) \left[ \frac{l(l+1)}{\rho^2(r)} + \frac{1}{\rho(r)}\frac{d}{dr}\left(f(r)\frac{d\rho(r)}{dr}\right) \right] \,.
\end{equation}
Using the explicit form $\rho(r)=\sqrt{r^2+a^2}$, the potential simplifies to:
\begin{equation}
    V_{\text{eff}}(r) = f(r) \left[ \frac{l(l+1)}{r^2+a^2} + \frac{r f^{\prime}(r)}{r^2+a^2} + \frac{a^2 f(r)}{(r^2+a^2)^2} \right] \,.
\end{equation}
To determine the QNM frequencies and assess stability, we utilize the Pöschl-Teller approximation. We approximate the effective potential near its maximum $r^*_0$ with the function:
\begin{equation}
    V_{\text{PT}}(r^{*}) = \frac{V_{0}}{\cosh^{2}[\alpha(r^{*} - r^{*}_{0})]} \,,
\end{equation}
where $V_0 = V_{\text{eff}}(r^*_0)$ is the peak height, and $\alpha$ is the curvature parameter defined by:
\begin{equation}
    \alpha^2 = -\frac{1}{2V_{0}} \frac{d^2 V_{\text{eff}}}{dr^{*2}}\bigg|_{r^{*}=r^{*}_{0}} \,.
\end{equation}
The corresponding quasi-normal frequencies are given analytically by:
\begin{equation}
    \omega_n = \pm \sqrt{V_{0} - \frac{\alpha^2}{4}} - i\alpha\left(n + \frac{1}{2}\right) \,, \quad n=0,1,2,\dots
\end{equation}
We evaluated the potential parameters for the fundamental mode ($n=0$) using the metric parameters $G=1, M=1$, and $a=0.1$. This choice satisfies the non-extremality condition $GM > 2a/(3\pi)$. The maximum of the potential is located approximately at $r \approx 3GM$. The results are summarized in Table~\ref{tab:QNM_values}.

\begin{table}[h]
    \centering
    \setlength{\tabcolsep}{12pt}
    \renewcommand{\arraystretch}{1.2}
    \begin{tabular}{cccc}
        \hline\hline
        $l$ & $V_{0}$ & $\alpha$ & $V_{0} - \frac{\alpha^2}{4}$ \\ 
        \hline
        1 & 0.098697 & 0.184230 & 0.090212 \\ 
        2 & 0.246746 & 0.189166 & 0.237800 \\ 
        3 & 0.468821 & 0.190698 & 0.468490 \\ 
        \hline\hline
    \end{tabular}
    \caption{Parameters of the Pöschl-Teller potential fit for different multipole numbers $l$.}
    \label{tab:QNM_values}
\end{table}

As shown in Table~\ref{tab:QNM_values}, the quantity $V_{0} - \alpha^2/4$ is positive for all $l$, ensuring a real oscillation frequency. More importantly, the imaginary part of the frequency is:
\begin{equation}
    \text{Im}(\omega) = -\alpha\left(n + \frac{1}{2}\right) \,.
\end{equation}
Since $\alpha$ is real and positive, $\text{Im}(\omega) < 0$ for all modes. The time dependence of the perturbation is $\Psi \sim e^{-i\omega t} = e^{-i\text{Re}(\omega)t} e^{\text{Im}(\omega)t} = e^{-i\text{Re}(\omega)t} e^{-|\text{Im}(\omega)|t}$. Consequently, the perturbation undergoes damped oscillations and decays exponentially with time, confirming the linear mode stability of the black hole solution.

%

\end{document}